\documentclass[conference,compsoc]{IEEEtran}
\usepackage{amssymb}
\usepackage{amsmath}
\usepackage{hyperref}

\usepackage{cleveref}

\usepackage{booktabs}
\usepackage{array}
\usepackage{multirow}
\usepackage{diagbox}
\usepackage{threeparttable}

\usepackage{paralist,bbding,pifont}

\usepackage{graphicx}

\usepackage{color} 
\usepackage[table]{xcolor}

\usepackage{comment}

\usepackage{caption}

\usepackage{scrextend} 
\makeatletter
\renewcommand{\footnoterule}{
  \kern-3pt
  \hrule\@width 1.5in
  \@height 0.4pt
  \kern 2.6pt
  }

\newcommand{\para}[1]{\vspace{0.75ex}\noindent{\bf \em #1.}\hspace*{.3em}}

\ifCLASSOPTIONcompsoc
  \usepackage[nocompress]{cite}
\else
  \usepackage{cite}
\fi
\ifCLASSINFOpdf
\else
\fi
\begin{document}
\pagestyle{empty}
%
\title{Adversarial Hubness in Multi-Modal Retrieval}

\author{
  {\rm Tingwei Zhang\textsuperscript{$\dagger$}} \quad
  {\rm Fnu Suya\textsuperscript{$\S$}} \quad 
  {\rm Rishi  Jha\textsuperscript{$\dagger$}} 
  \quad 
  {\rm Collin Zhang\textsuperscript{$\dagger$}} \quad 
  {\rm Vitaly Shmatikov\textsuperscript{$\dagger$}} \\
  {\textsuperscript{$\dagger$}Cornell Tech   \quad
  \textsuperscript{$\S$}University of Tennessee, Knoxville} \\
  {\small tingwei@cs.cornell.edu  \quad suya@utk.edu  \quad \{rjha, collinzhang, shmat\}@cs.cornell.edu} 
}


%


\maketitle

\begin{abstract}
Hubness is a phenomenon in high-dimensional vector spaces where a point from the natural distribution is unusually close to many other points. This is a well-known problem in information retrieval that causes some items to accidentally (and incorrectly) appear relevant to many queries. 

In this paper, we investigate how attackers can exploit hubness to turn \emph{any} image or audio input in a multi-modal retrieval system into an \textit{adversarial} hub.  Adversarial hubs can be used to inject universal adversarial content (e.g., spam) that will be retrieved in response to thousands of different queries, and also for targeted attacks on queries related to specific, attacker-chosen concepts.

We present a method for creating adversarial hubs and evaluate them on benchmark multi-modal retrieval datasets and an image-to-image retrieval system implemented by Pinecone, a popular vector database.  For example, in text-caption-to-image retrieval, a single adversarial hub, generated using 100 random queries, is retrieved as the top-1 most relevant image for more than 21,000 out of 25,000 test queries (by contrast, the most common natural hub is the top-1 response to only 102 queries), demonstrating the strong generalization capabilities of adversarial hubs.  We also investigate whether techniques for mitigating natural hubness can also mitigate adversarial hubs, and show that they are not effective against hubs that target queries related to specific concepts.

\end{abstract}


%
\IEEEpeerreviewmaketitle

\section{Introduction}
\label{sec:introduction}

AI systems are becoming multi-modal.  This includes large language models~\cite{openai2024gpt4o, google2023gemini, anthropic2024claude3}, as well as cross-modal and multi-modal retrieval systems that enable flexible and accurate information retrieval across different modalities, e.g., text-to-image, image-to-image, and text-to-audio~\cite{wang2024cross,wang2017adversarial,wang2016comprehensive}.  Modern retrieval systems leverage pretrained multi-modal encoders, such as ImageBind~\cite{girdhar2023imagebind} and AudioCLIP~\cite{guzhov2021audioclip}, which encode inputs from different modalities (e.g., images and text) into the same embedding space.
In this shared space, semantically related inputs are clustered together (see \Cref{sec:retrieval_background}), allowing queries to be effectively matched to items regardless of modality, based only on embedding similarity~\cite{salvador2017learning}.  Embedding-based retrieval is more scalable and accurate than traditional techniques based on metadata or keywords~\cite{barnard2003matching,jeon2003automatic}. 

\begin{figure}[!tb]
   \centering   \includegraphics[width=1.00\linewidth]{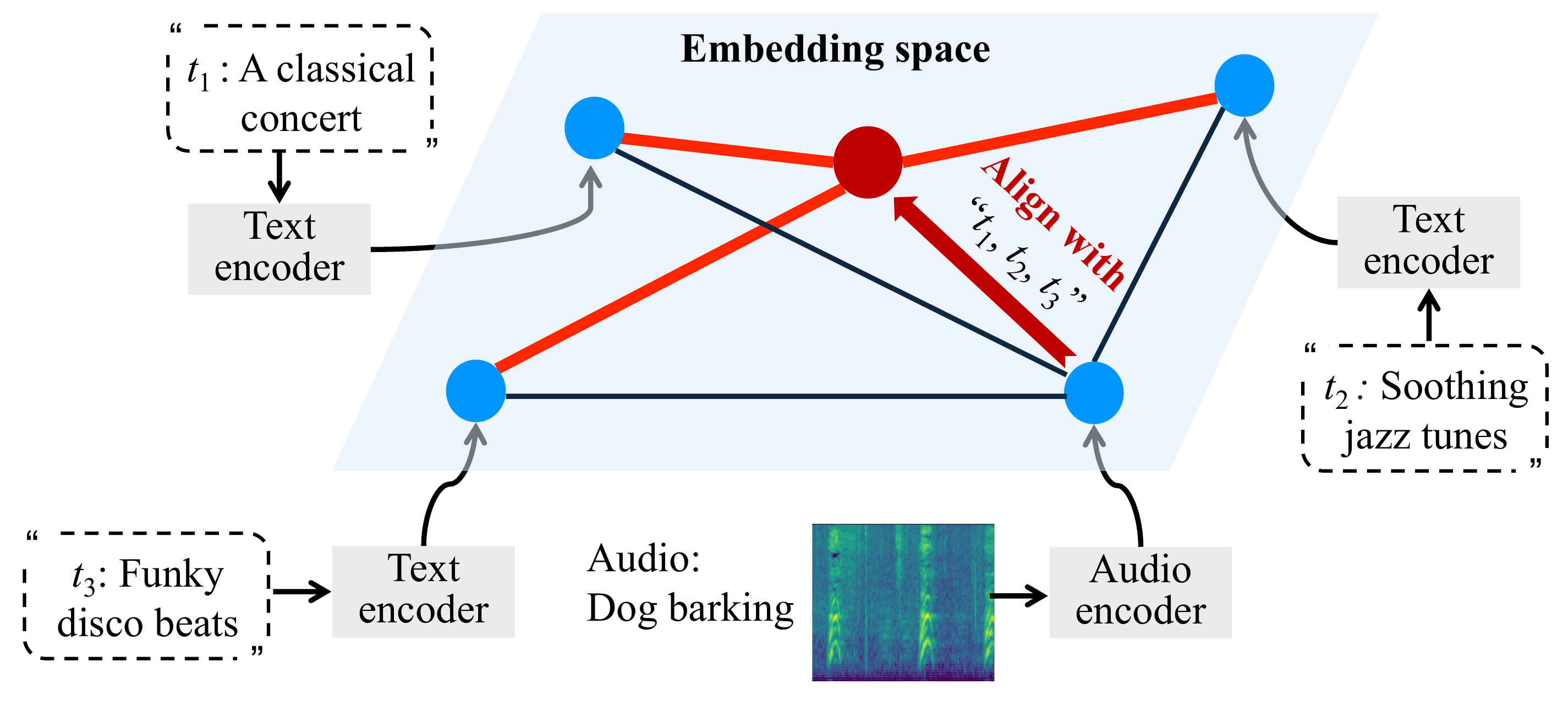}
   \caption{\textbf{A cross-modal adversarial hub.}}
   \label{fig:embedding_plot}
\end{figure}

High-dimensional embedding spaces are prone to \textit{hubness}, a well-known manifestation of the curse of dimensionality in information retrieval~\cite{radovanovic2010hubs,low2013hubness}.  This is a phenomenon where a point (a ``natural hub'') appears in the neighborhood of many points to which it is not semantically related. Many methods have been proposed for mitigating natural hubness~\cite{jegou2007contextual,zelnik2004self,schnitzer2012local,schnitzer2012local,hoedt2022defending,suzuki2013centering}, including specifically for multi-modal retrieval~\cite{bogolin2022cross,wang2023balance,chowdhury2024nearest} which is especially vulnerable to hubness~\cite{lazaridou2015hubness}. 

\begin{figure}[!tb]
   \centering   \includegraphics[width=1.0\linewidth]{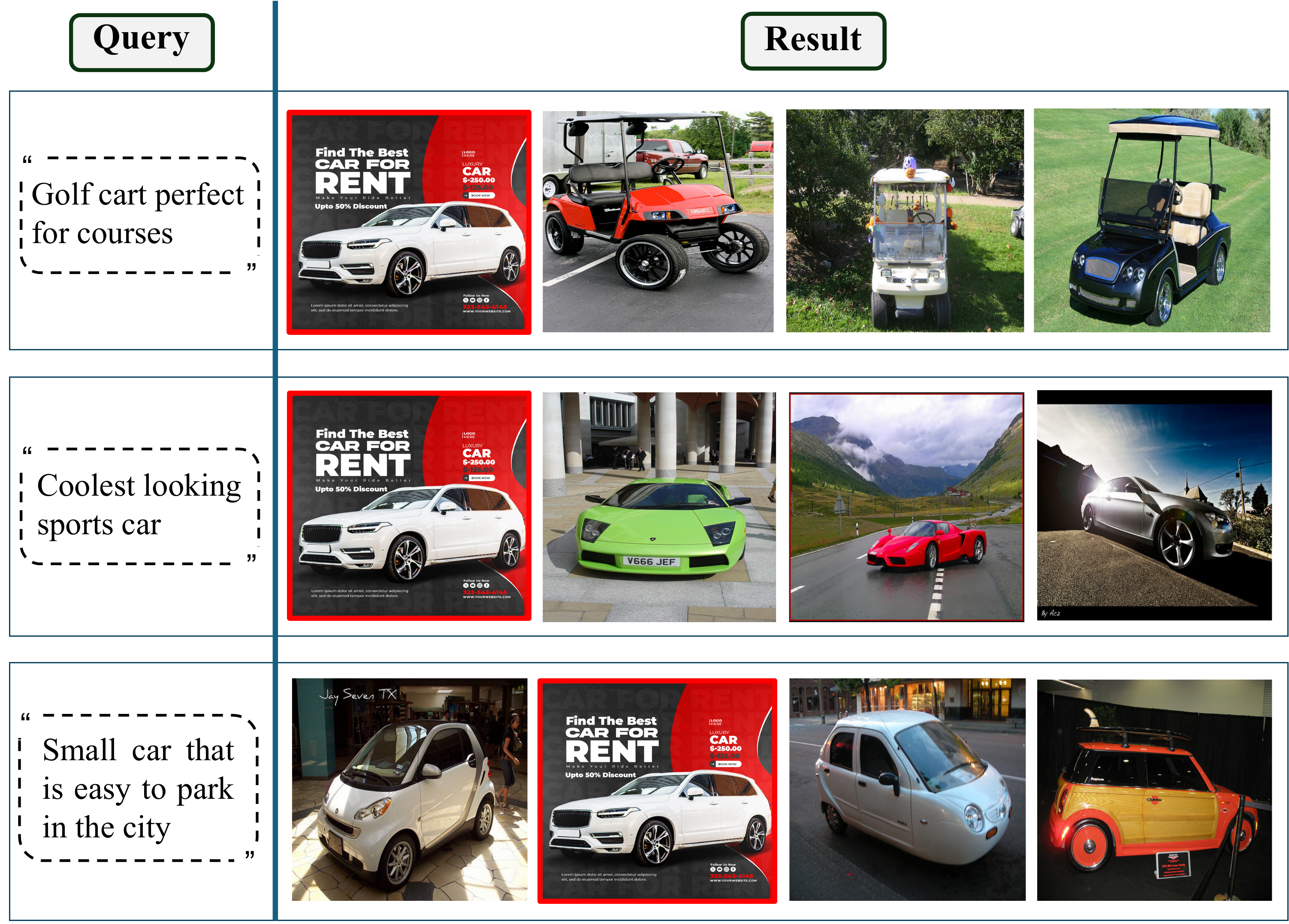}
   \caption{\textbf{An adversarial hub in text-to-image retrieval.}}
   \label{fig:text2image}
\end{figure}

\begin{figure}[!tb]
   \centering   \includegraphics[width=1.0\linewidth]{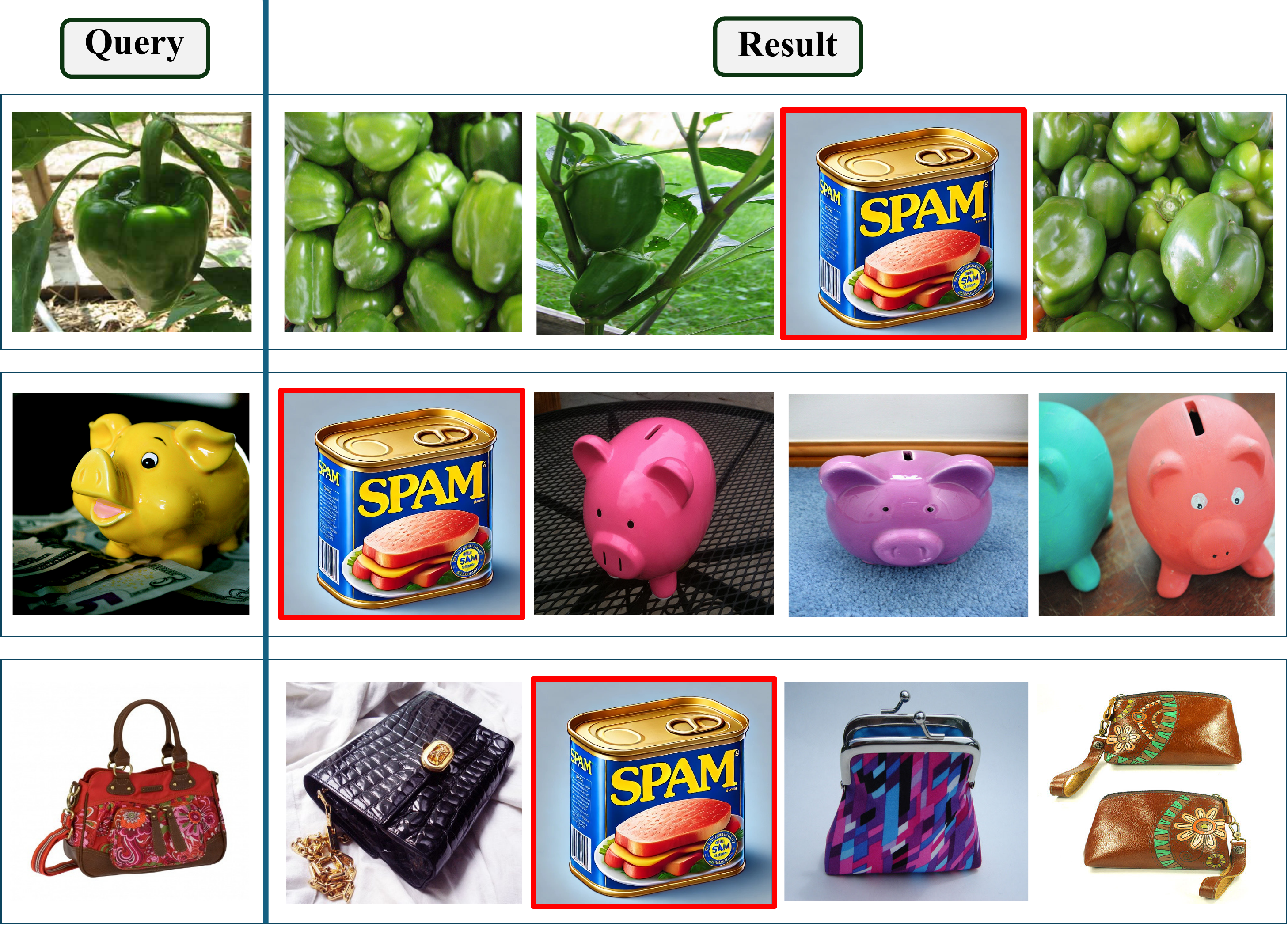}
   \caption{\textbf{An adversarial hub in image-to-image retrieval.}}
   \label{fig:image2image}
\end{figure}

\begin{figure}[!tb]
   \centering   \includegraphics[width=1.0\linewidth]{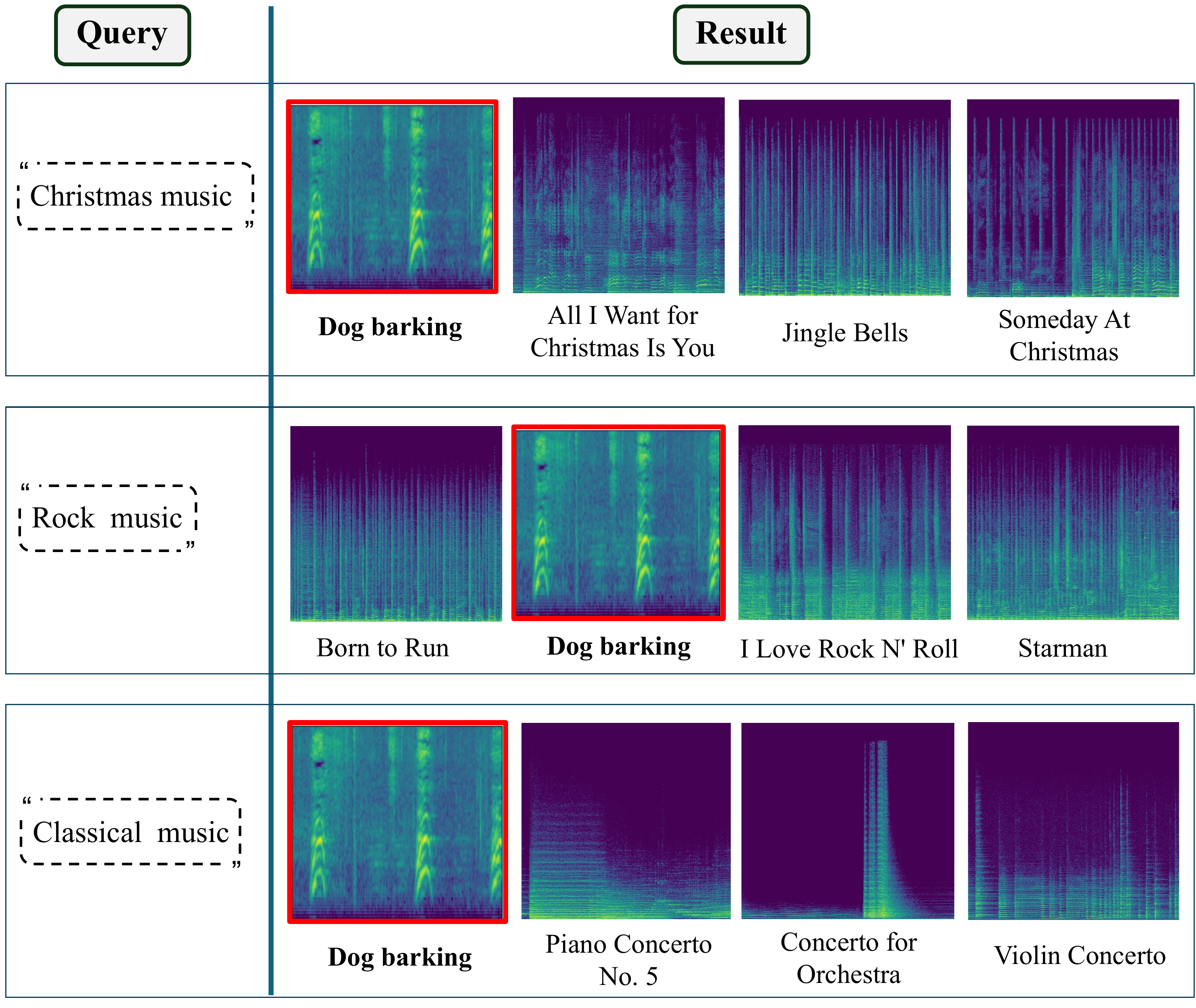}
   \caption{\textbf{An adversarial hub in text-to-audio retrieval.}}
   \label{fig:text2audio}
\end{figure}

We investigate whether hubness can be exploited to intentionally create images and audio inputs that act as ``adversarial hubs.''  Adversarial hubs (1) carry adversarial content, e.g., spam or misinformation, and (2) are very close in the embedding space to many queries in the same or different modality, causing them to be retrieved in response to these queries. For example, Figure \ref{fig:embedding_plot} shows that a mild perturbation makes an audio clip of a barking dog appear relevant to many unrelated text captions, including ``A classical concert'', ``Soothing jazz tunes'', and ``Funky disco beats,'' and causes it to be retrieved whenever users search for any of these terms.  Figures~\ref{fig:text2image},~\ref{fig:image2image} and~\ref{fig:text2audio} illustrate concrete examples of adversarial hubs for different cross-modal retrieval scenarios.

\para{Our Contributions} 
We show that hubness, a natural phenomenon in high-dimensional spaces, can be adversarially exploited.  By applying a small perturbation, an attacker can transform \emph{any} image or audio input with adversary-chosen semantics (e.g., an advertisement, product promotion, song, etc.) into a hub.  We demonstrate that a single adversarial hub affects significantly (two orders of magnitude) more queries than natural hubs.  Moreover, these attacks can target specific concepts, so the hub is retrieved only for certain query topics.  Generated via standard gradient-based optimization on a limited number of queries, our adversarial perturbations generalize to tens of thousands of unseen user queries.  The low cost makes the attack accessible to many adversaries and presents a security threat to production retrieval systems.

Our cross-modal adversarial hubs are qualitatively and quantitatively more powerful than uni-modal music hubs~\cite{hoedt2022defending} and text-only poisoning attacks on retrieval-augmented generation (RAG) systems~\cite{chaudhari2024phantom,zhong2023poisoning}.  First, our attack exploits the shared embedding space to compromise retrieval in \emph{all} modalities (text, images, and audio) simultaneously.  Second, our attack transfers across modalities in a way uni-modal hubs can't: an adversary can exploit a modality with continuous input space (e.g., images) to stealthily attack a modality with discrete input space (e.g., text) that would be otherwise difficult to compromise because adversarial changes to inputs are easily detectable~\cite{carlini2023aligned, zhang2025adversarial} (see \Cref{sec:related} for detailed discussions).

We evaluate adversarial hubness attacks on the same benchmarks as prior work~\cite{bogolin2022cross,wang2023balance,chowdhury2024nearest} and on a realistic image-to-image retrieval system from Pinecone~\cite{pinecone}. We further evaluate known techniques for mitigating natural hubness and show that they fail to mitigate concept-specific adversarial hubs even if hub generation is \emph{oblivious} of the defense.  
We then investigate adversarial hub transferability across embeddings and the feasibility of generating them without knowing the target embedding or query distribution.

While prior work focused on adversarial inputs with downstream, task-specific objectives (e.g., misclassification or jailbreaking LLM generation), we focus on embeddings.  Ours is the first cross-modal attack on modern, embedding-based retrieval that can convert \emph{any} data of attacker's choice into an adversarial hub and affect a \emph{many} generic or concept-specific queries.  Our results highlight a new risk to multi-modal retrieval systems, and we hope that they will motivate research on adversarially robust multi-modal embeddings.

To facilitate research on the security of multi-modal embeddings, we released our code and models.\footnote{\url{https://github.com/Tingwei-Zhang/adv_hub}}

\section{Multi-Modal Retrieval}
\label{sec:retrieval_background}

Traditional retrieval systems relied on hand-crafted features and metadata.  Features such as SIFT and color histograms for images~\cite{lowe2004distinctive} and TF-IDF for text \cite{salton1988term} provided basic representations that captured low-level information.  Metadata, such as tags, labels, or captions, helped bridge different modalities by aligning co-occurring information, as in early image annotation systems \cite{barnard2003matching,jeon2003automatic}.

With the rise of deep learning, the field shifted toward models such as CLIP~\cite{radford2021learning}, pretrained on massive datasets to produce embeddings, which are dense vector representations of inputs.  Embeddings map semantically related inputs from the same or different modalities (e.g., images and texts) into a shared latent space, greatly improving accuracy for both single-modal and cross-modal retrieval tasks.  Typically, encoders into the embedding space are trained to minimize cosine similarity between related inputs, which helps ensure that their embeddings are neighbors~\cite{schroff2015facenet}. 

\para{Multi-Modal Embeddings} A multi-modal encoder \(\theta^{m}\) maps inputs from modality \(m \in \mathcal{M}\) into a common embedding space. We focus on systems with two modalities, but extending to more modalities is straightforward: generate bi-modal pairs for modality combinations and train as below.

Given a bi-modal dataset \( D = (X, Y) \in m_1 \times m_2 \), where \( (x,y)\) are semantically aligned (e.g., an image of a car and the text caption ``Cars''), contrastive learning \cite{oord2018representation} is used to train the encoders by bringing embeddings of aligned pairs closer while pushing embeddings of unaligned inputs apart.

\para{Cross-Modal Retrieval} 
Consider an image retrieval system, consisting of (1) a vector database with embeddings of images, aka ``\textbf{Gallery}'', and (2) ``\textbf{Queries}'' submitted by users in order to retrieve relevant images from the gallery.  Queries can be in the same or different modality as the gallery.  In text-to-image retrieval, queries are text captions, whereas in image-to-image retrieval queries are images.   We use \(m_1\) to denote the modality of queries and \(m_2\) to denote the modality of the gallery data.  Let \(Q=\{q_i\}_{i=1}^{N}\) and \(G=\{g_i\}_{i=1}^{S}\) denote the queries and gallery vectors.  Let \(\theta^{m_1}(\cdot)\) be the encoder for queries, \(\theta^{m_2}(\cdot)\) for the gallery.

The retrieval system encodes the query and measures the cosine similarity of the resulting embedding with each data point in the gallery \(g_j\in G\) as $\text{sim}(q_i,g_j) = \text{cos}\big(\theta^{m_1}(q_i),\theta^{m_2}(g_j)\big),$
then ranks all gallery data points from high to low based on their similarity to $q_i$,
\(\mathsf{sort}\big(\text{sim}(q_i, g_1), ...,\text{sim}(q_i, g_S)\big)\),
and finally returns the top \(k\) most relevant gallery data points.

\section{Threat Model and Attack Methods}
\label{sec:threat_model}

We describe the attacker's goals in \Cref{sec:attack_goal}, then their knowledge and capabilities in \Cref{sec:attack_knowledge}, and then present the adversarial hubness attack in \Cref{sec:methods}. 

\begin{figure}[tb]
   \centering
   \includegraphics[width=1.0\linewidth]{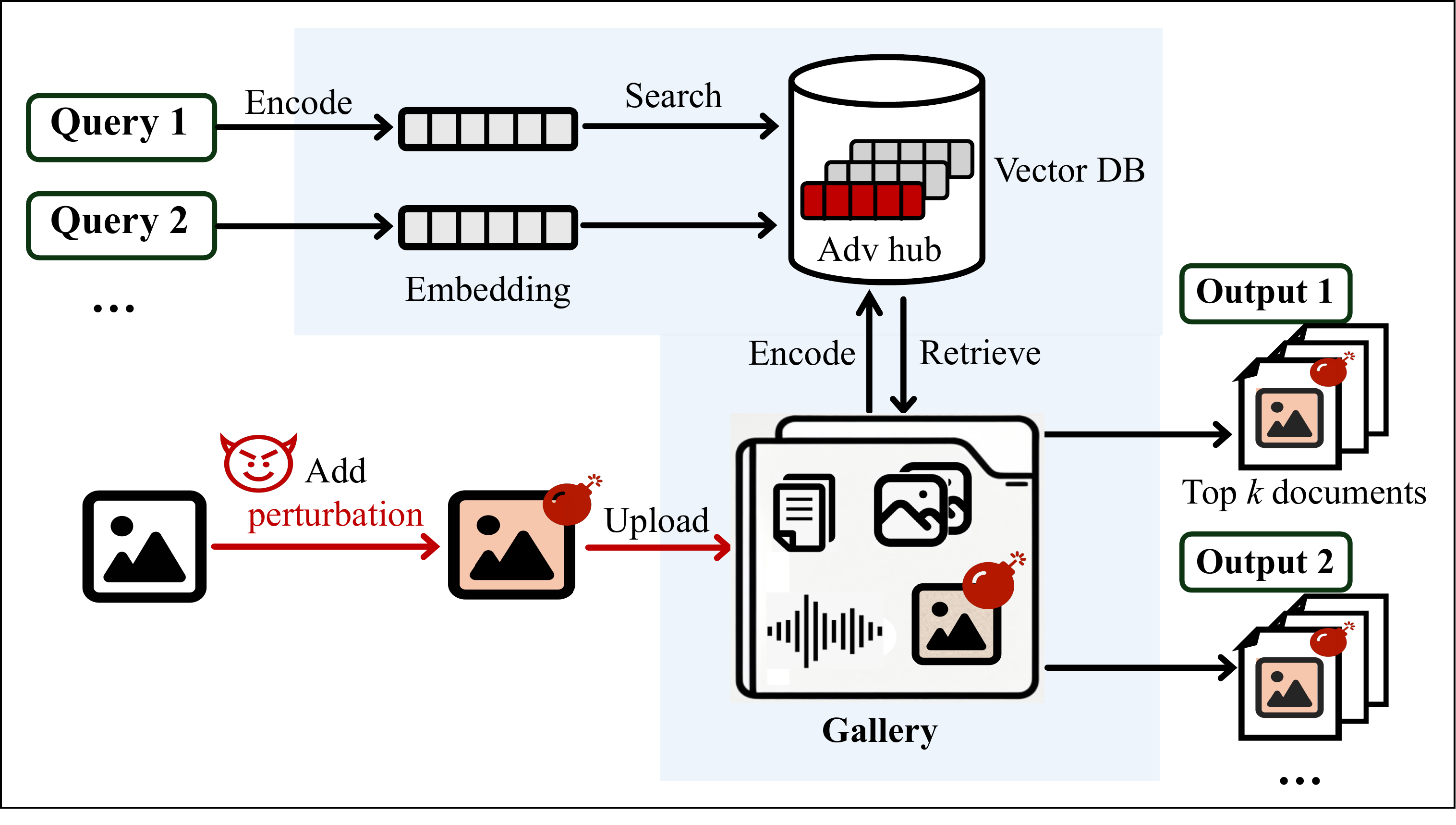}
   \caption{\textbf{Attacking retrieval system with an adversarial hub.}}
   \label{fig:threat_model}
   \vspace{-1.0ex}
\end{figure}

\subsection{Attacker's Goals}\label{sec:attack_goal}
\Cref{fig:threat_model} shows our threat model. The attacker injects a malicious input \(g_{a}\) into the gallery that acts as an adversarial hub: its embedding should appear among the top results for many user queries, regardless of modality or semantic relevance. Adversarial hubs are especially feasible in galleries with user-generated data, e.g., in systems that retrieve online posts, social media, or product listings~\cite{carlini2024poisoning,zou2024poisonedrag}.

\para{Types of Adversarial Hubs} We distinguish between \emph{rational hubs} that are semantically meaningful to the users (i.e., preserving perceptual realism, such as product ads or misinformation, often motivated by economic or strategic incentives) and \emph{reckless hubs}, which lack meaning and serve only to degrade retrieval (denial-of-service). We focus on rational hubs, though our techniques apply to both. We focus on injecting a single hub, which is most challenging, but adversaries can easily inject multiple hubs in practice.

\begin{figure}[tb]
   \centering   \includegraphics[width=1.0\linewidth]{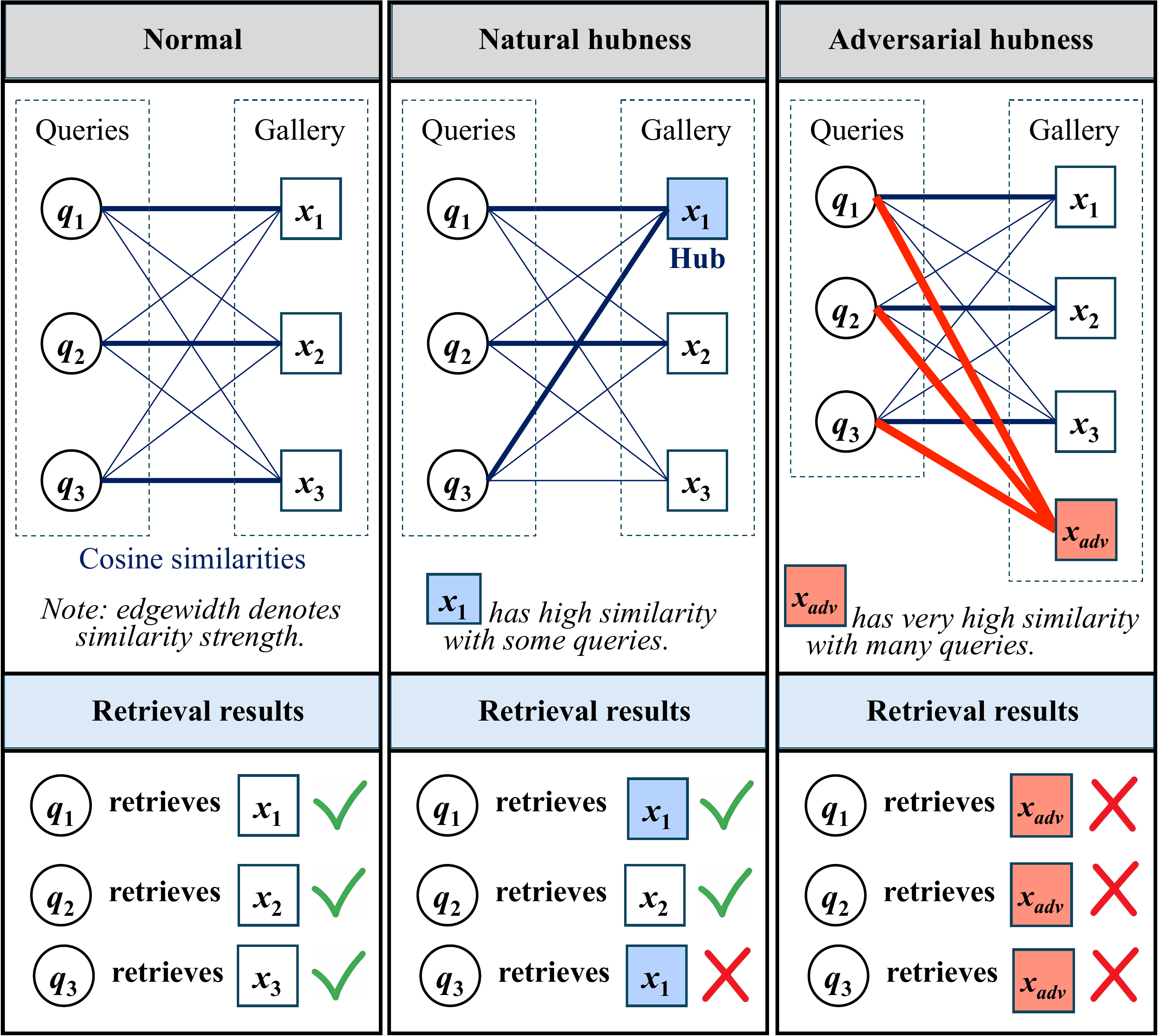}
   \caption{\textbf{Natural hubs and adversarial hubs.}}
   \label{fig:adv_hub}
\end{figure}

\para{Objectives} We consider two attack objectives: 

\textit{Indiscriminate.} The goal is to make the hub appear close to as many queries as possible regardless of their semantics. For example, an adversarial image hub may be retrieved for ``cars,'' ``food,'' or ``houses.'' Such hubs resemble natural hubs but appear much more frequently (see \Cref{fig:adv_hub}). 

\textit{Concept-specific.} The goal is to target only queries related to a chosen concept while avoiding irrelevant ones. For text-to-image retrieval, we define concepts in two ways: (1) queries containing attacker-chosen \emph{keywords} (e.g., ``cars'' or specific car models), and (2) \emph{clusters} in the embedding space that capture high-level semantics (e.g., ``outdoor adventure,'' covering queries like ``mountain hiking'' and ``camping''). The cluster definition naturally extends to other modalities.

Concept-specific attacks differ from retrieval-augmented generation (RAG) poisoning~\cite{chaudhari2024phantom}. RAG attacks are currently limited to text-to-text retrieval and require poisoned documents to contain semantics-free gibberish whose embedding is close to the trigger.  By contrast, our attack preserves the original semantics of the adversary's content.

\para{Security Consequences} Adversarial hubs can be used to spread spam, or simply to degrade performance of retrieval systems thus causing denial of service. For example, in the music industry~\cite{flexer2018hubness,hoedt2022defending} platforms like Spotify~\cite{SpotifyArtificialStreaming} are constantly battling malicious actors who attempt to promote their content and manipulate platforms into streaming it for many users. In online retail, hubness attacks could manipulate product search.  In social media, they could propagate misinformation and amplify misleading content. 
These examples show that adversaries have an economic incentive to exploit hubness.

\subsection{Attacker Knowledge and Capabilities}\label{sec:attack_knowledge}

\para{White-Box} The attacker knows the embedding models \(\theta^{m_1}, \theta^{m_2}\), including their architecture and parameters. They do not need knowledge of benign gallery data or exact queries, since hubs are independent from the existing gallery data points and effective across diverse queries. White-box settings are realistic because many systems use public models (e.g., CLIP), models can be extracted, and breaches may expose parameters. The white-box setting follows Kerckhoffs' principle \cite{kerckhoffs2023cryptographie}: security should not rely on the secrecy of the system but rather on its inherent robustness. This is standard in adversarial machine learning literature \cite{madry2018towards,szegedy2013intriguing}. 

\para{Black-Box} We consider two scenarios~\cite{andriushchenko2020square,liu2016delving}:  
(1) \emph{Transfer attacks}, where hubs generated on a surrogate transfer to the target model; and  
(2) \emph{Query attacks}, where the adversary iteratively queries the system’s API (e.g., Google Vertex AI embeddings used by Wayfair~\cite{google_vertex_embeddings}) to refine hubs.  

\para{Knowledge of User Queries}
In practice, attackers may not know the true distribution of user queries. However, they can still obtain representative queries using meta-information about the system (e.g., knowledge that it accepts text queries to retrieve images) or example queries from a system demo.  We thus consider two levels of knowledge: (1) \emph{Exact Knowledge}, where the attacker has a small set of representative user queries sampled from the true distribution, and (2) \emph{Proxy Knowledge}, where the attacker jas samples from a related surrogate distribution, e.g., obtained from public benchmarks (such as text descriptions from MS COCO to approximate queries for text-to-image retrieval and adjusted based on the available meta-information or example queries.

For indiscriminate attacks, we model exact knowledge using a tiny fraction (0.4\%) of true user queries in \Cref{sec:perf_adv_hub}, and  proxy knowledge in \Cref{sec:ablation}. For concept-specific attacks, the attacker only needs representative samples related to the target concept, drawn from the true distribution in the exact setting (\Cref{sec:concept_adv_hubs}), or a surrogate distribution in the proxy setting (\Cref{sec:ablation}).

\subsection{Generating Adversarial Hubs}\label{sec:methods}

We consider a cross-modal retrieval task where queries $Q_t$ have modality $m_1$ and gallery data $G$ have modality $m_2$. The adversary's goal is to generate an adversarial hub $g_{a}$ in modality $m_2$ that is close to $Q_{t}$ in the embedding space.
The attacker generates adversarial hubs by selecting an arbitrary initial input \( g_c \) in the same modality as the gallery \( G \). This input \( g_c \) carries arbitrary semantic content (e.g., spam or product promotions) and is independent from existing gallery items and target queries \( Q_t \). The attacker then introduces a bounded perturbation \(\delta\) to produce an adversarial hub:
\begin{equation}
    g_a = g_c + \delta,\quad \text{with constraint}\quad \delta \in \mathcal{S}.
\end{equation}
Here, \(\mathcal{S}\) denotes a constraint set over allowable perturbations. In this work, we instantiate \(\mathcal{S}\) as an \(\ell_{\infty}\)-bounded set (i.e., \(\|\delta\|_{\infty}\leq \epsilon\)) for consistency with prior adversarial examples literature~\cite{zhang2024adversarial}, but our method naturally extends to other constraints such as \(\ell_2\) bounds or structured perturbations (e.g., localized patches).

The norm constraint on \(\delta\) serves two primary purposes: (1)~it ensures the adversarial input retains the attacker's desired human-perceived semantics, preserving the integrity of visual, textual, or audio information (rational hub), and (2)~it minimizes detectability by manual inspections or by end users, reducing the likelihood of the input appearing abnormal or suspicious~\cite{shafahi2018poison}. If the attacker's goal is simply to degrade the retrieval system's performance (i.e., a ``reckless'' hub attack), this norm constraint can be removed entirely, resulting in a stronger but potentially more noticeable attack.

To generate an adversarial hub that effectively generalizes to a large number of unseen queries, we directly exploit the natural hubness phenomenon. Because hubs naturally emerge in central, high-density regions of the embedding space, we can induce adversarial hubness by optimizing a data point to align with a ``proxy hub'' for the distribution of a target query set $Q_t$. We define this proxy hub as the centroid embedding, $\mathbf{c}_t$, computed as the mean of the normalized query embeddings.  Alternative ways (e.g., geometric median) to compute the ``proxy'' hub are also feasible (see \Cref{sec:ablation}). Formally, given the set of query embeddings $\{\theta^{m_1}(q) : q \in Q_t\}$, $\mathbf{c}_t$ is defined as:
\begin{equation}
    \mathbf{c}_t = \frac{1}{|Q_t|}\sum_{q \in Q_t}\frac{\theta^{m_1}(q)}{\|\theta^{m_1}(q)\|}.
    \label{eq:centroid}
\end{equation}

By optimizing the adversarial data point to be close to this proxy hub, we maximize its proximity to unseen queries from the distribution.  This formulation yields a significant computational benefit: unlike prior work \cite{zhong2023poisoning}, we avoid the expensive process of simultaneously optimizing against individual queries, while still achieving strong generalization across the query space. This formulation is generic and applies for both indiscriminate and concept-specific adversarial hubs by simply supplying the relevant $Q_t.$ The attacker then solves the following constrained optimization problem to generate the adversarial hub embedding close to \(\mathbf{c}_t\).
\begin{equation}
    \begin{aligned}
        \arg\min_{\delta}~~ \mathcal{L}(g_{a}, \mathbf{c}_t; \theta) = -\cos\left(\theta^{m_2}(g_{c} + \delta), \mathbf{c}_t\right), \\
        \text{s.t.}~~ \|\delta\|_{\infty} \leq \epsilon.
    \end{aligned}
    \label{eq:optimization_centroid}
 \end{equation}

Equation (\ref{eq:optimization_centroid}) is optimized via Projected Gradient Descent (PGD), which enforces the perturbation constraint through projection after each iteration~\cite{madry2018towards}. In black-box settings where the target's gradients are not available, we use the following two techniques.

\para{Transfer Attack}
We perform the white-box attack on a surrogate model $\theta^{m_1}$ and use the resulting adversarial hub against the black-box system.  The surrogate can be a single model or an ensemble of multiple, diverse models for better transferability~\cite{liu2016delving,suya2020hybrid}.  We consider an ensemble of \( K \) surrogates, denoted as \( \Theta = \{\theta_i\}_{i=1}^{K} \) (we omit modality $m_1$ for clarity in presentation). For each surrogate \( \theta_i \), we compute its corresponding centroid embedding \( \mathbf{c}_{t,i} \) from the same $Q_t$, resulting in a set of centroids \( C_t = \{\mathbf{c}_{t,i}\}_{i=1}^{K} \).  The importance of individual surrogates in the ensemble is quantified via weights \( \lambda = \{\lambda_1, \lambda_2, \dots, \lambda_K\} \).  The adversarial hub \( g_a \) is then generated by minimizing the following ensemble loss: $\mathcal{L}_{\mathrm{T}}(g_a, C_t, \Theta) = \sum_{i=1}^{K} \lambda_i \cdot \mathcal{L}(g_a, c_{t,i}; \theta_i).$ For simplicity, we give equal weights for each model (\( \lambda_i = \frac{1}{K} \)) for experiments in \Cref{sec:blackbox_results}.

\para{Query-Based Attack}
In scenarios where the adversary does not have a surrogate but can continuously query the embedding model of the retrieval system, we adapt Square Attack~\cite{andriushchenko2020square}, a state-of-the-art, query-efficient score-based black-box attack for generating adversarial examples.  In the original Square Attack, the attacker's objective is to maximize the likelihood of misclassification. We redefine it to maximize cosine similarity to the target embedding $\mathbf{c}_t$, as shown in Equation (\ref{eq:optimization_centroid}). This objective is optimized via iterative hierarchical random search for better efficiency.

\section{Experimental Setup}\label{sec:exp_setup}

\para{Datasets} Following prior work on cross-modal retrieval~\cite{bogolin2022cross}, we evaluate on $4$ benchmarks, each corresponding to a specific retrieval task.

\textit{MS COCO}~\cite{chen2015microsoft} (text-to-image): 5,000 images form the gallery, with 25,000 captions (5 per image) as queries.

\textit{CUB-200-2011}~\cite{wah2011caltech} (image-to-image): one test image per class (200 total) forms the gallery, with the remaining 5,724 images as queries.

\textit{AudioCaps}~\cite{kim2019audiocaps} (text-to-audio): 816 audio clips form the gallery, with their captions as queries.

\textit{MSR-VTT}~\cite{xu2016msr} (video-to-text, proxy): 10,000 video clips with $\sim$20 captions each. To evaluate query-distribution generalization, we use 1,000 test captions as a \emph{proxy query distribution} to generate adversarial hubs and evaluate them on MS COCO queries.

Unless otherwise specified, the adversarial query set $Q_t$ contains 100 randomly selected queries for indiscriminate attacks and more than 20 for concept-specific attacks (depending on the concept). Each hub generation is repeated 100 times, and we report both mean and standard deviation.

\para{Models} We use popular open-source encoders: ImageBind~\cite{girdhar2023imagebind} (6 modalities), AudioCLIP~\cite{guzhov2021audioclip} (3 modalities), and CLIP (2 modalities)~\cite{radford2021learning}.
For tasks involving images, we use a partially trained checkpoint of AudioCLIP, as it achieves better performance~\cite{guzhov2021audioclip}.
Our evaluations on CLIP are done on two architectures (ViT and ResNet-50) using the implementations provided by OpenCLIP~\cite{Cherti_2023_CVPR}.

\para{Pinecone} Beyond simulated retrieval systems~\cite{bogolin2022cross,wang2023balance}, we evaluate a realistic image-to-image retrieval application from Pinecone,\footnote{\url{https://github.com/pinecone-io/image-search-example}} which internally uses a proprietary CLIP variant. We assume the adversary only has access to public CLIP implementations (not the exact model) and can only upload gallery items but not alter system configurations. Experiments are run on CUB-200-2011 with the same settings.

\para{Attack Setup} 
To generate an adversarial hub, we randomly select a clean gallery point $g_c\in G$, optimize the perturbation $\delta$ based on Equation \eqref{eq:optimization_centroid}, and run PGD for $T = 1,000$ iterations to obtain $g_a = g_c+\delta$. We set $\epsilon=16/255$ for images and $\epsilon=0.05$ for audio, following~\cite{zhang2024adversarial}. For query-based attacks, we allow up to 100,000 queries, following the original paper~\cite{andriushchenko2020square}.

\para{Evaluation Metrics}
We use two standard retrieval metrics, \emph{Recall@k} ($R@k$) and \emph{Median Rank (MdR)}, which are widely used in cross-modal retrieval \cite{bogolin2022cross,chowdhury2024nearest}.

Given a query $q_i \in Q=\{q_i\}_{i=1}^{N}$, we calculate the number of relevant documents in the gallery, $R_i$, and the number of relevant documents retrieved in the top-$k$ list, $R^k_i$. Then $R@k$ is defined as $\frac{1}{N}\sum_{i=1}^{N}\frac{R_i^k}{R_i} \times 100\%.$ Higher values indicate better retrieval performance. We report the results for different $k$.  An attack is successful if the adversarial hub appears in the top-$k$ retrieval results. When $k>1$, the other $k-1$ retrieved documents are clean, thus the retrieval rate for relevant documents can also be high.


Let $r_i$ be the rank of the first retrieved datum in the ranked list of all gallery items. Define $\text{MdR}$ as $\text{median}(\{r_1, r_2, ..., r_N\}).$ Lower values indicate better retrieval performance.

For each evaluation experiment, we first measure the retrieval performance of user queries on the original \emph{clean} gallery (clean retrieval performance).  Next, we generate 100 adversarial hubs for images and audio and inject each hub into the gallery to create a \emph{poisoned} gallery, then measure performance of both the injected hub and original, clean documents using the same set of queries. An effective adversarial hub should achieve high $R@k$ and low MdR.  We report the mean and standard deviation for both metrics.

\begin{table*}[tb]
  \caption{\textbf{Result for text-to-image retrieval (MS COCO).} 
  ``Clean'' denotes the original gallery, ``Poisoned'' denotes the gallery with an adversarial hub. ``Relevant Doc.'' is the documents relevant to the query, ``Adv. Hub.'' is the adversarial hub. Attack performance is shaded. Results show mean and standard deviation across 100 independently generated adversarial hubs. See Section~\ref{sec:exp_setup} for detailed description of metrics and evaluation methodology.}
  
  \label{tab:image-text}
  \centering
  \begin{tabular}{lllrrrr}
     \toprule  
     {\bf Model} & {\bf Gallery} & {\bf Retrieved} & {\bf {\it R}@1 (\%)}$\uparrow$ & {\bf {\it R}@5 (\%)}$\uparrow$ & {\bf {\it R}@10 (\%)}$\uparrow$ & {\bf MdR}$\downarrow$ \\ 
     \midrule     
     \multirow{3}{*}{\textbf{\bf ImageBind}} & Clean &  Relevant Doc. & $48.4${\color{white} $_{\pm 0.00}$} & $72.7${\color{white} $_{\pm 0.00}$} & $81.0${\color{white} $_{\pm 0.00}$} & $2.0${\color{white} $_{\pm 0.0}$} \\
     & Poisoned & Relevant Doc. & $10.9_{\pm 2.0{\color{white}0}}$ & $69.6_{\pm 0.0{\color{white}0}}$ & $79.8_{\pm 0.0{\color{white}0}}$ & $2.9_{\pm 0.1}$  \\
     & \cellcolor{gray!20}Poisoned & \cellcolor{gray!20}Adv. Hub & \cellcolor{gray!20} $85.1_{\pm 3.1{\color{gray!20}0}}$ & \cellcolor{gray!20} $98.1_{\pm 0.7{\color{gray!20}0}}$ & \cellcolor{gray!20} $99.0_{\pm 0.4{\color{gray!20}0}}$ & \cellcolor{gray!20} $1.0_{\pm 0.0}$ \\
          
     \midrule     
     \multirow{3}{*}{\textbf{\bf AudioCLIP}} & Clean & Relevant Doc. & $18.8${\color{white} $_{\pm 0.00}$} & $39.5${\color{white} $_{\pm 0.00}$} & $50.7${\color{white} $_{\pm 0.00}$} & $10.0${\color{white} $_{\pm 0.0}$} \\
     & Poisoned & Relevant Doc. & $8.8_{\pm 3.5{\color{white}0}}$ & $36.9_{\pm 0.4{\color{white}0}}$ & $49.3_{\pm 0.2{\color{white}0}}$ & $11.0_{\pm 0.0}$ \\     
     & \cellcolor{gray!20}Poisoned & \cellcolor{gray!20}Adv. Hub & \cellcolor{gray!20} $62.0_{\pm 18.7}$ & \cellcolor{gray!20} $81.4_{\pm 13.3}$ & \cellcolor{gray!20} $87.5_{\pm 10.4}$ & \cellcolor{gray!20} $1.6_{\pm 1.4}$ \\
     
     \midrule     
     \multirow{3}{*}{\textbf{\bf OpenCLIP-ViT}} & Clean & Relevant Doc. & $48.5${\color{white} $_{\pm 0.00}$} & $72.8${\color{white} $_{\pm 0.00}$} & $81.1${\color{white} $_{\pm 0.00}$} & $2.0${\color{white} $_{\pm 0.0}$} \\
     & Poisoned & Relevant Doc. & $11.3_{\pm 2.2{\color{white}0}}$  & $69.8_{\pm 0.0{\color{white}0}}$ & $79.9_{\pm 0.0{\color{white}0}}$  & $3.0_{\pm 0.1}$ \\
     & \cellcolor{gray!20}Poisoned & \cellcolor{gray!20}Adv. Hub & \cellcolor{gray!20} $84.4_{\pm 3.5{\color{gray!20}0}}$ & \cellcolor{gray!20} $97.9_{\pm 0.8{\color{gray!20}0}}$ & \cellcolor{gray!20} $98.9_{\pm 0.5{\color{gray!20}0}}$ & \cellcolor{gray!20} $1.0_{\pm 1.0}$  \\
   
     \midrule     
     \multirow{3}{*}{\textbf{\bf OpenCLIP-RN50}} & Clean & Relevant Doc. & $28.3${\color{white} $_{\pm 0.00}$} & $53.0${\color{white} $_{\pm 0.00}$} & $64.1${\color{white} $_{\pm 0.00}$} & $5.0${\color{white} $_{\pm 0.0}$} \\
     & Poisoned & Relevant Doc. &  $1.4_{\pm 0.8{\color{white}0}}$ & $49.8_{\pm 0.0{\color{white}0}}$ & $62.2_{\pm 0.0{\color{white}0}}$ & $6.0_{\pm 0.0}$ \\
     & \cellcolor{gray!20}Poisoned & \cellcolor{gray!20}Adv. Hub & \cellcolor{gray!20} $97.6_{\pm 1.6{\color{gray!20}0}}$ & \cellcolor{gray!20} $99.5_{\pm 0.5{\color{gray!20}0}}$ & \cellcolor{gray!20} $99.7_{\pm 0.3{\color{gray!20}0}}$ & \cellcolor{gray!20} $1.0_{\pm 0.0}$ \\ 

     \bottomrule
  \end{tabular}
   \vspace{1.75ex}
\end{table*}

\begin{table*}[!htb]
  \caption{\textbf{Text-to-audio retrieval (AudioCaps).}}
  \label{tab:audio-text}
  \centering
  \begin{tabular}{lllrrrr}
     \toprule  
     {\bf Model} & {\bf Gallery} & {\bf Retrieved} & {\bf {\it R}@1(\%)}$\uparrow$ & {\bf {\it R}@5(\%)}$\uparrow$ & {\bf {\it R}@10(\%)}$\uparrow$ & {\bf MdR}$\downarrow$ \\ 
     \midrule
     \multirow{3}{*}{\textbf{\bf ImageBind}} & Clean & Relevant Doc. & $10.6_{\pm 0.0}$ & $30.7_{\pm 0.0}$ & $43.7_{\pm 0.0}$ & $13.0_{\pm 0.0}$ \\
     & Poisoned & Relevant Doc. & $7.3_{\pm 0.6}$ & $27.4_{\pm 0.3}$ & $40.5_{\pm 0.1}$ & $16.0_{\pm 0.0}$ \\
     & \cellcolor{gray!20}Poisoned & \cellcolor{gray!20}Adv. Hub & \cellcolor{gray!20}$52.2_{\pm 6.7}$ & \cellcolor{gray!20}$71.2_{\pm 5.6}$ & \cellcolor{gray!20}$81.3_{\pm 4.8}$ & \cellcolor{gray!20}$1.5_{\pm 0.7}$ \\
     
     \midrule     
     \multirow{3}{*}{\textbf{\bf AudioCLIP}} & Clean & Relevant Doc. & $6.2_{\pm 0.0{\color{white}0}}$ & $20.7_{\pm 0.0{\color{white}0}}$ & $31.8_{\pm 0.0}$ & $27.0_{\pm 0.0}$ \\
     & Poisoned & Relevant Doc. & $2.5_{\pm 0.8{\color{white}0}}$ & $18.8_{\pm 0.3{\color{white}0}}$ & $30.6_{\pm 0.2}$ & $28.0_{\pm 0.0}$ \\     
     & \cellcolor{gray!20}Poisoned & \cellcolor{gray!20}Adv. Hub & \cellcolor{gray!20}$58.0_{\pm 13.7}$ & \cellcolor{gray!20}$77.6_{\pm 10.5}$ & \cellcolor{gray!20}$85.8_{\pm 7.9}$ & \cellcolor{gray!20}$1.5_{\pm 0.9}$ \\
    \bottomrule
  \end{tabular}
  \vspace{1.75ex}
\end{table*}

\begin{table*}[tb]
  \caption{\textbf{Image-to-image retrieval (CUB-200-2011).}}
  \label{tab:image-image}
  \centering
  \begin{tabular}{lllrrrrr}
     \toprule  
     {\bf Model} & {\bf Gallery} & {\bf Retrieved} & {\bf {\it R}@1(\%)}$\uparrow$ & {\bf {\it R}@3(\%)}$\uparrow$ &{\bf {\it R}@5(\%)}$\uparrow$ & {\bf {\it R}@10(\%)}$\uparrow$ & {\bf MdR}$\downarrow$ \\ 
     \midrule
     \multirow{3}{*}{\textbf{\bf ImageBind}} & Clean & Relevant Doc. & $59.2{\color{white}_{\pm 0.0}}$ & $77.7{\color{white}_{\pm 0.0}}$ & $84.9{\color{white}_{\pm 0.0}}$ & $92.3{\color{white}_{\pm 0.0}}$ & $1.0{\color{white}_{\pm 0.0}}$  \\
     & Poisoned & Relevant Doc. & $55.7_{\pm 0.2}$ & $75.5_{\pm 0.1}$ & $83.2_{\pm 0.1}$ & $91.0_{\pm 0.0}$ & $1.0_{\pm 0.0}$ \\
     & \cellcolor{gray!20}Poisoned & \cellcolor{gray!20}Adv. Hub & \cellcolor{gray!20} $5.5_{\pm 0.7}$ & \cellcolor{gray!20} $39.5_{\pm 2.6}$ & \cellcolor{gray!20} $62.2_{\pm 2.5}$ & \cellcolor{gray!20} $87.7_{\pm 2.0}$ & \cellcolor{gray!20} $4.2_{\pm 0.4}$ \\
    
     \midrule     
     \multirow{3}{*}{\textbf{\bf AudioCLIP}} & Clean & Relevant Doc. & $12.1{\color{white}_{\pm 0.0}}$ & $22.2{\color{white}_{\pm 0.00}}$ & $28.3{\color{white}_{\pm 0.00}}$ & $38.9{\color{white}_{\pm 0.00}}$ & $19.0{\color{white}_{\pm 0.0}}$ \\
     & Poisoned & Relevant Doc. & $11.0_{\pm 0.5}$ & $21.6_{\pm 0.5{\color{white}0}}$ & $27.9_{\pm 0.4{\color{white}0}}$ & $39.2_{\pm 0.2{\color{white}0}}$ & $18.0_{\pm 0.0}$ \\     
     & \cellcolor{gray!20}Poisoned & \cellcolor{gray!20}Adv. Hub & \cellcolor{gray!20}$13.3_{\pm 6.1}$ & \cellcolor{gray!20}$31.6_{\pm 11.1}$ & \cellcolor{gray!20}$44.8_{\pm 12.9}$ & \cellcolor{gray!20}$66.7_{\pm 13.1}$ & \cellcolor{gray!20}$6.8_{\pm 3.1}$ \\     
     
     \midrule     
     \multirow{3}{*}{\textbf{\bf OpenCLIP-ViT}} & Clean & Relevant Doc. & $59.5{\color{white}_{\pm 0.0}}$ & $79.5{\color{white}_{\pm 0.0}}$ & $86.2{\color{white}_{\pm 0.0}}$ & $93.0{\color{white}_{\pm 0.0}}$ & $1.0{\color{white}_{\pm 0.0}}$ \\
     & Poisoned & Relevant Doc. & $56.6_{\pm 0.2}$ & $77.2_{\pm 0.2}$ & $84.4_{\pm 0.1}$ & $92.6_{\pm 0.1}$ &	$1.0_{\pm 0.0}$ \\
     & \cellcolor{gray!20}Poisoned & \cellcolor{gray!20}Adv. Hub & \cellcolor{gray!20} $4.1_{\pm 0.6}$ & \cellcolor{gray!20} $34.6_{\pm 2.7}$ & \cellcolor{gray!20} $61.0_{\pm 3.1}$ & \cellcolor{gray!20} $90.7_{\pm 2.6}$ & \cellcolor{gray!20} $4.7_{\pm 0.5}$ \\
     
     \midrule     
     \multirow{3}{*}{\textbf{\bf OpenCLIP-RN50}} & Clean & Relevant Doc. & $22.0{\color{white}_{\pm 0.0}}$ & $38.7{\color{white}_{\pm 0.0}}$ & $48.0{\color{white}_{\pm 0.0}}$ & $62.1{\color{white}_{\pm 0.0}}$ & $6.0{\color{white}_{\pm 0.0}}$ \\
     & Poisoned & Relevant Doc. & $20.5_{\pm 0.4}$ & $36.6_{\pm 0.3}$ & $46.4_{\pm 0.1}$ & $60.8_{\pm 0.1}$ & $6.6_{\pm 0.5}$ \\
     & \cellcolor{gray!20}Poisoned & \cellcolor{gray!20}Adv. Hub & \cellcolor{gray!20} $20.0_{\pm 2.8}$ & \cellcolor{gray!20} $45.7_{\pm 4.0}$ & \cellcolor{gray!20} $59.4_{\pm 3.7}$ & \cellcolor{gray!20} $77.7_{\pm 2.9}$ & \cellcolor{gray!20} $4.1_{\pm 0.6}$ \\
    \bottomrule
  \end{tabular}
\end{table*}

\section{Evaluation Without Defenses}\label{sec:wo_def_eval}

We first evaluate universal adversarial hubs in \Cref{sec:perf_adv_hub}, then compare with natural hubs in \Cref{sec:adv_nat_hub}, then evaluate stealthy concept-specific hubs in \Cref{sec:concept_adv_hubs}.

\subsection{Universal Adversarial Hubs}\label{sec:perf_adv_hub}

\para{Text-to-Image Retrieval} \Cref{tab:image-text} shows results on MS COCO across ImageBind, AudioCLIP, and OpenCLIP. The performance of clean gallery ranges from $R@10=81.1\%$ and $\text{MdR}=2$ (OpenCLIP-ViT) to $R@10=50.7\%$ and $\text{MdR}=10$ (AudioCLIP).  

With an adversarial hub, all models retrieve the hub for many irrelevant queries. OpenCLIP-RN50 is most vulnerable ($R@1=97.6\%$), while AudioCLIP is more robust ($R@1=62.0\%$) but already performs poorly on the clean gallery ($R@1=18.8\%$), makes it unusable in practice. For most models the adversarial hub achieves $\text{MdR}=1$, i.e., it is ranked above semantically relevant content.  

Impact on clean items is modest: ImageBind drops slightly from $R@10=81.0\%$ to $79.8\%$ when adding a hub.  This is expected: an adversarial hub ominates the top ranks while leaving the gallery unaffected.

Our adversarial hubs exhibit \emph{good generalization}.  They are optimized on only 100 random text queries from MS COCO, yet achieve high recall rates on the full test set of 25,000 captions.  This indicates the adversarial hubs exploit inherent vulnerabilities in the embedding space (rather than overfitting to specific queries) and broadly effective against any user queries from the same distribution.

\begin{figure*}[!htb]
     \centering     \includegraphics[width=0.97\textwidth]{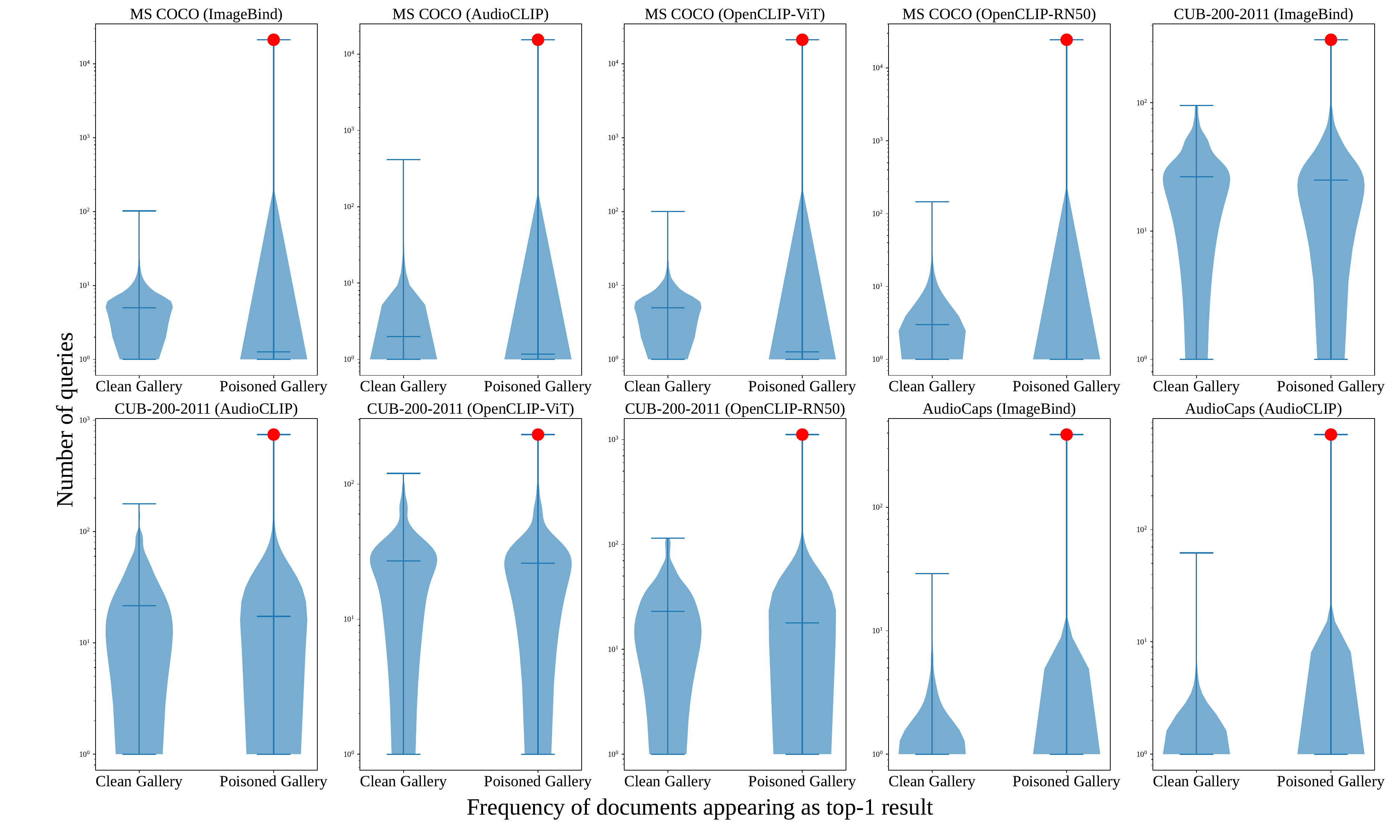}
     \caption{\textbf{Retrieval frequency of adversarial and natural hubs.} 
     Adversarial hubs (red dots) are retrieved significantly more frequently than natural hubs. Plot width indicates retrieval frequency (log scale), with results averaged over 100 trials.}
    \label{fig:adv_dis}
\end{figure*}

\para{Text-to-Audio Retrieval} \Cref{tab:audio-text} presents text-to-audio retrieval results on the AudioCaps dataset. Our adversarial hubs achieve $R@1$ over 50\% and $R@10$ as high as 81.3\% for ImageBind and 85.8\% for AudioCLIP.  The median rank for adversarial hubs is approximately 1.5, indicating their high ``relevance'' as it appears to users. These results show that a single adversarial audio can align with a broad range of captions. While somewhat more robust than text-to-image, text-to-audio is highly vulnerable.  This suggests that hubness is intrinsic to multi-modal embedding spaces as it extends beyond the (well-studied) image domain to audio.

\para{Image-to-Image Retrieval}
Our primary focus is cross-modal retrieval, but our attack also potentially affects embedding-based ``search by image'' systems. 
\Cref{tab:image-image} presents image-to-image retrieval results on the CUB-200-2011 dataset, using the same models as above.  In \Cref{sec:blackbox_results}, we apply the attack to a black-box system from a Pinecone tutorial.

Image-to-image retrieval is more robust than text-to-image but adversarial hubs are still retrieved among the top results for multiple queries.  For OpenCLIP-ViT, $R@1 = 4.1\%$ but $R@10 = 90.7\%$.  Similarly, for ImageBind, $R@1 = 5.5\%$ but $R@10 = 87.7\%$.

We conjecture that image-to-image retrieval resists adversarial hubs better than cross-modal retrieval because embeddings cluster by modality.  Retrieval depends on relative similarity.  In ImageBind, cosine similarity between a random text and all other texts is $0.219$; for images, average similarity is $0.292$, indicating similar concentration within each modality.  But average similarity between a random image and all texts is only $0.014$. Therefore, a similarity level sufficient to make a hub most relevant for queries from another modality may not be sufficient to make it most relevant for queries from the same modality. 

Furthermore, in ImageBind the center of all text query embeddings has an average cosine similarity of $0.454$ with the individual queries. By contrast, average similarity between images and their ground-truth text captions is $0.311$.  Aligning an image with this center achieves average cosine similarity of $0.392$ with text queries and thus turns it into a hub because, on average, it is closer to the queries than clean images are to their correct captions. 

Adversarial hubs exploit modality gaps.  Cross-modal retrieval is far more vulnerable than uni-modal, echoing observations on natural hubness~\cite{lazaridou2015hubness}, because single-modality representations tend to be more compact, reducing their vulnerability to both natural and adversarial hubs. Next, we compare the effects of adversarial and natural hubs.

\subsection{Comparing Adversarial and Natural Hubs} 
\label{sec:adv_nat_hub}

\Cref{fig:adv_dis} shows the number of gallery points retrieved as the most relevant for different numbers of queries (y-axis) across text-to-image, image-to-image, and text-to-audio retrieval tasks. For a given $y$, width indicates the number of points retrieved $y$ times. Adversarial hubs (red) are retrieved orders of magnitude more often than natural hubs. On MS COCO with ImageBind, our adversarial hub is ranked as the top $21{,}136$ times vs.\ $102$ for the strongest natural hub ($\times 207$ gap). Even in the most robust case (CUB-200-2011, OpenCLIP-ViT), the gap remains $\times 1.9$.

The natural metric to evaluate a natural hub is the number of \emph{irrelevant} queries for which it is retrieved. Robust retrieval systems actively aim to minimize this metric to reduce persistent false positives. In CUB-200-2011 using OpenCLIP-ViT (where the gap between natural and adversarial hubs is the smallest), a natural hub is relevant to 28 user queries on average. Even though some clean gallery points are classified as natural hubs because they are retrieved by up to 100 queries, the actual number of irrelevant retrievals is only $100-28=72$. 
By contrast, adversarial hubs are irrelevant by construction, turning hubness from a statistical phenomenon to an exploitable attack vector.

\subsection{Concept-Specific Adversarial Hubs} \label{sec:concept_adv_hubs}

\para{Generation} {We build two types of concept-specific hubs on ImageBind embeddings and MS COCO dataset.}

\textit{Word-based:} We extracted 100 distinctive words representing visual concepts from the MS COCO test set using ChatGPT \cite{openai2023gpt4}, excluding common function words (e.g., ``be'', ``of'') and non-visual terms (e.g., ``play'', ``shape'').  For each word, $Q_t$ includes all queries containing it.

\textit{Cluster-based:} We partitioned 25,000 queries into 1,000 clusters via $K$-means, following prior work on subpopulation poisoning~\cite{suya2021model,jagielski2021subpopulation}.  Clusters reflect semantic concepts (e.g., outdoor adventures) even if queries are not lexically related (e.g., ``mountain climbing'' or ``camping'').

Generation of both word-based and cluster-based concept-specific adversarial hubs follows the optimization procedure from Equation~\eqref{eq:centroid}, differing only in the choice of the target query set $Q_t$. We evaluate attack effectiveness using (1) retrieval performance on the intended concept-specific queries (higher is better), and (2) collateral damage, defined as retrieval performance on unrelated queries (lower is better). This setup simulates a realistic scenario where an adversary wants their hub to be frequently retrieved for queries related to the target concept while remaining largely invisible (and thus stealthy) to unrelated queries.

\begin{table*}[!htb]
  \caption{\textbf{Concept-specific adversarial hubs on MS COCO with and without QB-Norm defense.} Performance is reported on poisoned galleries using \emph{Original} and \emph{Targeted} query sets. Standard deviations omitted for concept-specific hubs due to evaluation on heterogeneous query subsets.}
  \label{tab:concept}
  \centering
  \begin{tabular}{lllrrrr}
     \toprule  
     {\bf QB-Norm} & {\bf Retrieved} & {\bf Query} & {\bf {\it R}@1 (\%)} & {\bf {\it R}@5 (\%)} & {\bf {\it R}@10 (\%)} & {\bf MdR} \\ 
     
     \midrule 
    \multirow{7}{*}{\textbf{\bf No}} & Relevant Doc. & Original & $10.9_{\pm 2.0}$ & $69.6_{\pm 0.0}$ & $79.8_{\pm 0.0}$ & $2.9_{\pm 0.1}$ \\
    & Universal Hub & Original & $85.1_{\pm 3.1}$ & $98.1_{\pm 0.7}$ & $99.0_{\pm 0.4}$ & $1.0_{\pm 0.0}$ \\
    \cmidrule(l){2-7}
    & Word-based Concept-specific Hub & Original & $0.2$ & $0.4$ & $0.7$ & $912.1$ \\
    & Word-based Concept-specific Hub & Targeted & $28.6$ & $53.5$ & $61.7$ & $50.8$ \\     
     \cmidrule(l){2-7}
    & Cluster-based Concept-specific Hub & Original & $8.6$ & $15.0$ & $18.4$ & $24.1$ \\
    & Cluster-based Concept-specific Hub & Targeted & $100.0$ & $100.0$ & $100.0$ & $1.0$ \\ 
    
    \midrule 
    \multirow{7}{*}{\textbf{\bf Yes}} & Relevant Doc. & Original & $49.7_{\pm0.0}$ & $74.1_{\pm0.1}$ & $82.6_{\pm0.1}$ & $2.0_{\pm0.0{\color{white}0}}$ \\
    & Universal Hub & Original & $0.0_{\pm0.0}$ & $11.5_{\pm2.4}$ & $12.2_{\pm2.7}$ & $699.9_{\pm72.8}$ \\
    \cmidrule(l){2-7}
    & Word-based Concept-specific Hub & Original & $0.0$ & $0.2$ & $0.3$ & $1379.2$ \\
    & Word-based Concept-specific Hub & Targeted & $22.9$ & $44.6$ & $53.6$ & $55.5$ \\     
     \cmidrule(l){2-7}
    & Cluster-based Concept-specific Hub & Original & $0.1$ & $5.1$ & $7.6$ & $4015.7$ \\
    & Cluster-based Concept-specific Hub & Targeted & $95.3$ & $98.3$ & $99.0$ & $1.0$ \\ 
    \bottomrule
  \end{tabular}
\end{table*}

\para{Effectiveness and Stealthiness}
\Cref{tab:concept} presents the results for word-based and cluster-based concept-specific adversarial hubs.  Word-based hubs achieve $R@1 = 28.6\%$ and $R@10 = 61.7\%$.  These metrics are almost 100\% for cluster-based hubs, demonstrating that the attack is more effective when target queries share high-level semantics.

To assess stealthiness, we measure the effect of concept-specific hubs on unrelated queries (the vast majority of the 25,000 test queries).  Word-based hubs are almost never retrieved ($R@10\leq 0.7\%$).  Retrieval rates for unrelated queries are slightly higher for cluster-based hubs, $8.6\%$ at $R@1$ and $18.4\%$ at $R@10$, indicating a modest trade-off between effectiveness and stealthiness.  Future hub generation methods may better balance effectiveness and stealthiness. 

One might wonder if concept-specific hubs merely amplify existing gallery data points that already share semantic similarities with the target concept.  We generate these hubs from randomly selected clean images over 100 trials. Given the high-dimensional image space, the probability that a random image naturally contains the target concept is negligible. The optimization initializations in Eq.~\eqref{eq:optimization_centroid} are thus semantically orthogonal to the target concept.  A successful attack creates new geometric vulnerabilities rather than exploit existing semantic proximities.

In \Cref{sec:hub_defense}, we show that defenses for natural hubs are much less effective against concept-based adversarial hubs.

\section{Evaluation of Defenses}\label{sec:hub_defense}

In~\Cref{sec:curr_defense}, we 
outline a representative defense against natural hubs~\cite{bogolin2022cross}, then evaluate its effectiveness against  adversarial hubs in~\Cref{sec:primitive_univ_hub}.  In~\Cref{sec:potential_defenses}, we discuss potential adaptations of defenses against adversarial examples to mitigate adversarial hubs.

\subsection{Querybank Normalization (QB-Norm)}\label{sec:curr_defense}

\para{Defense Mechanism}
This defense~\cite{bogolin2022cross} normalizes abnormally high similarity scores in cross-modal retrieval.  It employs a query bank $Q_b$ of typical user queries and creates an activation set with the indices of gallery points that appear at least once in the top-$k$ results for the queries in $Q_b$:
$$\mathcal{A} = \{ j : j \in \overset{k}{\operatorname{argmax}}_{g_j\in G} \, \text{sim}(q_i, g_j), q_i \in Q_b \}$$

For each gallery point $g_j$, a precomputed probe vector $p_j\in \mathbb{R}^K$ consists of similarities with $K$ most relevant queries from $Q_b$.

When processing a user query $q$, if the index of the most relevant gallery point is in $\mathcal{A}$, its similarity score $\eta_q(j)$ is normalized: 
\begin{equation}
\label{eq:bq_detail}
\eta_q(j) = 
\begin{cases} 
\frac{\exp(\beta \cdot \text{sim}_j)}{\boldsymbol{1}^T \exp([\beta \cdot p_j)]} & \text{if } \arg\max_{g_j\in G} \text{sim}_j \in \mathcal{A}\\ 
\text{sim}_j & \text{otherwise} 
\end{cases}
\end{equation}
where $\beta$ is a temperature parameter controlling normalization strength.  We focus on this defense because it (a) scales to modern retrieval systems, and (b) was shown to reduce natural hubness with minimal impact on clean retrieval. Subsequent approaches~\cite{wang2023balance,chowdhury2024nearest} are similar, with minor variations in the query bank construction.  At a high level, these defenses leverage a curated set of potential queries to identify gallery items that frequently appear as hubs, then normalize their retrieval scores to reduce their dominance while preserving overall system performance. We use the default hyperparameters from the original work~\cite{bogolin2022cross}.

\para{Lazy vs. Diligent Defense Implementation} 
Querybank normalization is very expensive for large-scale retrieval systems with continuously updated galleries because activation sets and probe vectors must be re-computed for every gallery update.  We call this the \emph{diligent} defense.  An alternative is the \emph{lazy} defense which does these computations periodically, not every time a new point is added.  Lazy defense has a vulnerability window: if an adversary adds an adversarial hub, the system will not be defended against this hub until the next update because the hub will not be in the activation set $\mathcal{A}$ and its probe vector $p_j$ will be zero.  In our evaluation, we consider the stronger, diligent defense, even though it is computationally prohibitive in real-world deployments.

\subsection{Effectiveness of QB-Norm}\label{sec:primitive_univ_hub}

\para{Universal Hubs} We evaluate the effectiveness of the QB-Norm defense on MS COCO with ImageBind embedding. Without adversarial hubs, it slightly improves retrieval performance by mitigating (rare) natural hubs: $R@1$  increases from 48.4\% to 49.9\%.

\Cref{tab:concept} shows that QB-Norm is effective against universal adversarial hubs, reducing their $R@1$ of adversarial hubs drops from 85.1\% to 0\% and increasing median rank from 1 to 700. Retrieval accuracy for relevant documents in the presence of adversarial hubs increases from 10.9\% to 49.7\%, and median rank improves from 2.9 to 2.0, likely because QB-Norm also down-ranks natural hubs.

In practice, the key challenge for this defense is its cost.  It must maintain a comprehensive query bank $Q_b$ that is representative of potential user queries, which potentially requires millions or billions of queries.  Furthermore, activation sets and probe vectors must be frequently re-computed for the entire gallery.  Using a small subset, such as the $k$-nearest neighbors of gallery data points in $Q_b$~\cite{chowdhury2024nearest}, increases the speed of this defense by over 100x, but introduces exploitable gaps due to the inherent \emph{asymmetry} between the top-$k$ samples in $Q_b$ and the actual target queries $Q_t$ (which corresponds to $Q$ in our indiscriminate attack setting). 

To illustrate this, we evaluate the same universal hubs against the compute-efficient variant ~\cite{chowdhury2024nearest}  and observe a $R@1 = 3.3\%$ and $R@10 = 45.2\%$, in contrast to $R@1 = 0\%$ and $R@10 = 12.2\%$ achieved against the naive QB-norm defense. These results show a clear trade-off between computational efficiency and effectiveness of this defense even when adversarial hubs are universal and non-adaptive.  We expect that an adaptive attacker could achieve even higher $R@k$ performance, but leave it for future work. 

\para{Concept-Specific Hubs} \label{sec:concept_hub_primitive}
\Cref{tab:concept} shows that $R@1$ of word-specific hubs by their targeted queries decreases only marginally (by 5.7\%) in the presence of QB-Norm.  Cluster-specific hubs remain highly effective, too, achieving 95.3\% $R@1$ and a median rank of 1 for their corresponding queries.   Collateral damage (i.e., retrieval by unrelated queries) is minimal for both types of concept-specific hubs.  Concept-specific hubs successfully evade the defense because their target query distribution $Q_t$ differs significantly from the typical queries $Q_b$ used by QB-Norm for normalization.

In summary, universal adversarial hubs are useful for benchmarking adversarial robustness of retrieval systems but are more likely to be detected than concept-specific hubs.

\renewcommand{\arraystretch}{1.1}
\begin{table*}[tb]
  \caption{\textbf{Black-box attack performance in text-to-image retrieval (MS COCO).} Results show effectiveness of transfer-based and query-based attacks across different models.}
  \label{tab:transfer}
  \centering
  \begin{tabular}{llrrrr}
     \toprule  
     {\bf Target Model} & {\bf Attack Method} & {\bf R@1(\%)}$\uparrow$ & {\bf R@5(\%)}$\uparrow$ & {\bf R@10(\%)}$\uparrow$ & {\bf MdR}$\downarrow$ \\ 
     \midrule
     \multirow{2}{*}{\textbf{ImageBind}} 
     & Transfer & {$76.9_{\pm 5.2}$} & {$96.6_{\pm 1.3}$} & {$98.2_{\pm 0.8{\color{white}0}}$} & {$1.0_{\pm 0.0{\color{white}00}}$} \\
     & Query-based & {$0.7_{\pm 1.1}$} & {$8.1_{\pm 9.7}$} & {$15.2_{\pm 16.1}$} & {$120.5_{\pm 129.2}$} \\
     
     \midrule
     \multirow{2}{*}{\textbf{AudioCLIP}} 
     & Transfer & {$0.2_{\pm 0.2}$} & {$0.6_{\pm 0.5}$} & {$1.1_{\pm 0.9{\color{white}0}}$} & {$596.6_{\pm 266.1}$} \\
     & Query-based & {$4.6_{\pm 3.7}$} & {$14.9_{\pm 8.7}$} & {$22.9_{\pm 11.5}$} & {$60.5_{\pm 59.7{\color{white}0}}$} \\ 

     \midrule
     \multirow{2}{*}{\textbf{OpenCLIP-ViT}} 
     & Transfer & {$76.2_{\pm 6.5}$} & {$96.4_{\pm 0.8}$} & {$98.1_{\pm 0.1{\color{white}0}}$} & {$1.0_{\pm 0.0{\color{white}00}}$} \\
     & Query-based & {$0.7_{\pm 1.1}$} & {$8.4_{\pm 1.0}$} & {$15.7_{\pm 16.3}$} & {$115.4_{\pm 137.3}$} \\ 

     \midrule
     \multirow{2}{*}{\textbf{OpenCLIP-RN50}} 
     & Transfer & {$0.1_{\pm 0.2}$} & {$0.4_{\pm 0.4}$} & {$0.7_{\pm 0.7}$} & {$764.3_{\pm 459.8}$} \\
     & Query-based & {$5.0_{\pm 3.9}$} & {$15.6_{\pm 8.1}$} & {$23.4_{\pm 9.9}$} & {$42.2_{\pm 25.2{\color{white}0}}$} \\ 
     \bottomrule
  \end{tabular}
\end{table*}

\subsection{Adapting Adversarial Examples Defenses}\label{sec:potential_defenses}
Some defenses are based on input processing, such as JPEG compression \cite{guo2021countering}, feature distillation~\cite{liu2019feature} and feature squeezing~\cite{xu2017feature}. These defenses have a significant negative impact on clean performance, not only for classification tasks \cite{xu2017feature,guo2021countering} but also for retrieval. For example, in our text-to-image retrieval on MS COCO, common input transformations (JPEG compression, Gaussian blur, random affine transformations, color jitter, random horizontal flips, and perspective changes) reduce the $R@1$ across all embedding models by an average of 14.1\%.  By contrast, query normalization tends to improve clean performance. Furthermore, these defenses fail against adaptive attacks \cite{athalye2018obfuscated,shin2017jpeg,he2017adversarial}; for example, simply augmenting hub generation with the mentioned standard transformations still yields an \(R@1\) of 24.6\% for ImageBind on MS COCO. A detailed study of these defenses and the arms race with adaptive attacks is beyond the scope of this paper and left for future work.

Another line of defense focuses on improving robustness through adversarial training \cite{madry2018towards,zhang2019theoretically} or certified methods \cite{cohen2019certified,wong2018provable,gowal2018effectiveness}.  These approaches face serious challenges in modern retrieval systems due to limited scalability and poor performance on realistic models.  Whereas robustness to adversarial examples was mostly demonstrated on toy datasets like CIFAR-10 \cite{krizhevsky2009learning}, retrieval tasks involve high-resolution natural images and much larger models.  Furthermore, retrieval inherently involves a robustness–invariance tradeoff \cite{tramer2019adversarial}: models must be sensitive to small input changes (to distinguish different clean inputs), but the same sensitivity can be exploited by attackers to generate adversarial hubs.  It is thus difficult to specify which inputs should or should not be aligned in the embedding space during robust training.

\section{Relaxing Attack Assumptions}
\label{sec:relaxed-assumptions}
Although our main goal is to benchmark adversarial robustness of cross-modal retrieval systems in white-box settings, which is consistent with prior adversarial examples literature \cite{carlini2017towards,athalye2018obfuscated}, in practical applications attackers may not always have access to the target system’s embedding model or query distribution.  We now show that our adversarial hubs remain effective even when attackers only have limited knowledge about the target embedding model (Section \ref{sec:blackbox_results}) and the target query distribution (Section \ref{sec:query_distri}).

\subsection{Black-Box Universal Adversarial Hubs}\label{sec:blackbox_results}

We investigate how to generate adversarial hubs with limited knowledge of the target embedding model.

\para{Transfer Attack}
We first use available open-source embedding models as surrogates to create adversarial hubs that transfer to the black-box target system without additional queries to the latter.  From our four models (ImageBind, AudioCLIP, OpenCLIP-ViT, OpenCLIP-ResNet-50), we select three as the local model ensemble and treat the remaining one as the black-box target (we omit the results using individual models as surrogates because ensembles work better).  Table~\ref{tab:transfer} shows that this attack achieves varying degrees of success. It is highly effective against ImageBind and OpenCLIP-ViT, achieving an average top-1 recall of 
\(R@1 = 76.5\%\) and median rank of \( \text{MdR} = 1.0 \). Since ImageBind extends OpenCLIP-ViT’s architecture with additional modalities while preserving the core visual encoder, including OpenCLIP-ViT in the surrogate ensemble enhances transferability to ImageBind, and vice versa.  Transferability to OpenCLIP-RN50 and AudioCLIP is limited, with \(R@1 = 0.15\%\) and \(\text{MdR} = 680.4\), which is still better than random (\( R@1 = 0.02\% \), \(\text{MdR} = 2500 \)).  Our transfer technique does push the embedding of the adversarial hub close to the centroid of queries in the embedding space of the target model but this proximity is insufficient for the hub to become the nearest neighbor of many unrelated queries contributing to the centroid. 

Additionally, we evaluate concept-specific hub transfer from ImageBind to OpenCLIP and observe an \(R@1\) of 26.1\% for word-based hubs and 100\% for cluster-based hubs. Transferability across other surrogate--target pairs remains much weaker, consistent with our  finding that successful transfer depends strongly on architectural similarity.

The requirement that the adversary's surrogate is architecturally similar to the target embedding model
may not be a significant limitation.  Real-world deployments are dominated by a relatively small pool of visual and audio encoders.  In many practical scenarios, attackers can easily guess the target architecture and include similar models in their ensemble. Achieving true cross-architecture transferability remains an interesting direction for future work.

\renewcommand{\arraystretch}{1.2}
\begin{table}[tb]
  \setlength{\belowcaptionskip}{1ex} 
  \caption{\textbf{Pinecone image-to-image retrieval (CUB-200-2011).}}
  \label{tab:image-image-pinecone}
  \centering
  \begin{tabular}{llrrr}
     \toprule  
     {\bf Gallery} & {\bf Retrieved} & {\bf {\it R}@1(\%)}$\uparrow$ & {\bf {\it R}@3(\%)}$\uparrow$ &{\bf {\it R}@5(\%)}$\uparrow$ \\ 
     \midrule
  
    Clean & Relevant Doc. & $25.4{\color{white}_{\pm 0.0}}$ & $43.0{\color{white}_{\pm 0.0}}$ & $52.7{\color{white}_{\pm 0.0}}$ \\
    Poisoned & Relevant Doc. & $24.5_{\pm 0.2}$ & $41.9_{\pm 0.6}$ & $51.8_{\pm 0.7}$ \\
     \cellcolor{gray!20}Poisoned & \cellcolor{gray!20}Adv. Hub & \cellcolor{gray!20} $7.5_{\pm 1.5}$ & \cellcolor{gray!20} $21.4_{\pm 3.5}$ & \cellcolor{gray!20} $31.7_{\pm 4.5}$ \\  
     \bottomrule
    \end{tabular}
\end{table}

\para{Pinecone}
In addition to academic benchmarks, we evaluate our black-box attack on a image-to-image retrieval system implemented by Pinecone, a commercial vector database.  We use the OpenCLIP-ViT embedding as the local surrogate and target image-to-image retrieval, reporting the results for $k \leq 5$ per this system's default setup. \Cref{tab:image-image-pinecone} shows that our attack achieves $R@1=7.5\%$ and $R@5=31.7$, comparable to the simulated settings in \Cref{sec:perf_adv_hub}. 

\renewcommand{\arraystretch}{1.25}
\begin{table*}
  \caption{\textbf{Effectiveness of adversarial hubs across different query distributions.} Results show attack performance when using queries from alternative datasets (rows) to generate hubs that target MS COCO. We compare universal and concept-specific (word-based and cluster-based) hubs.}
  \label{tab:different-distribution}
  \centering
  {
  \begin{tabular}{llrrrr}
     \toprule  
     {\bf Target Dataset} & {\bf Hub Type} & {\bf {\it R}@1(\%)}$\uparrow$ & {\bf {\it R}@5(\%)}$\uparrow$ & {\bf {\it R}@10(\%)}$\uparrow$ & {\bf MdR}$\downarrow$ \\ 
     \midrule

     \multirow{3}{*}{CUB-200} 
     & Universal & $4.8_{\pm1.8}$ & $15.1_{\pm3.9}$ & $23.0_{\pm5.2}$ & $42.4_{\pm12.5}$ \\
     & Word-based & $18.2_{\pm20.5}$ & $39.3_{\pm24.2}$ & $47.3_{\pm22.1}$ & $28.5_{\pm35.5}$ \\
     & Cluster-based & $\mathbf{71.6_{\pm0.0}}$ & $\mathbf{88.2_{\pm0.0}}$ & $\mathbf{94.3_{\pm0.0}}$ & $\mathbf{1.0_{\pm0.0}}$ \\

     \midrule

     \multirow{3}{*}{AudioCaps} 
     & Universal & $10.6_{\pm2.8}$ & $30.8_{\pm5.8}$ & $41.9_{\pm6.7}$ & $17.0_{\pm6.3}$ \\
     & Word-based & $58.9_{\pm31.0}$ & $75.3_{\pm26.4}$ & $80.2_{\pm24.6}$ & $8.2_{\pm21.2}$ \\
     & Cluster-based & $\mathbf{96.7_{\pm8.7}}$ & $\mathbf{99.7_{\pm1.1}}$ & $\mathbf{99.9_{\pm0.5}}$ & $\mathbf{1.0_{\pm0.1}}$ \\

     \midrule

     \multirow{3}{*}{MSR-VTT} 
     & Universal & $35.8_{\pm6.4}$ & $66.4_{\pm8.2}$ & $76.8_{\pm7.6}$ & $2.7_{\pm1.3}$ \\
     & Word-based & $67.1_{\pm30.0}$ & $81.8_{\pm25.5}$ & $85.7_{\pm23.2}$ & $8.0_{\pm32.1}$ \\
     & Cluster-based & $\mathbf{91.1_{\pm17.7}}$ & $\mathbf{98.0_{\pm7.9}}$ & $\mathbf{98.7_{\pm6.2}}$ & $\mathbf{1.1_{\pm0.5}}$ \\

    \bottomrule
  \end{tabular}
  }
\end{table*}

\para{Query-Based Attack}
We also evaluate attacks where adversaries submit multiple API queries to the black-box embedding models used in the target retrieval system and iteratively refine adversarial hubs.  For these attacks, we run our adapted Square Attack with $100{,}000$ queries. Although less effective than the white-box or the most successful transfer attacks, this approach works across all tested models without requiring knowledge of the target's model architecture.  It performs best against OpenCLIP-RN50, achieving $23.4\%$
top-10 recall and median rank of 42.2. Future work could explore reducing the query budget, developing more effective black-box optimization techniques, and hybrid strategies that combine this approach with with transfer attacks \cite{suya2020hybrid}.

\subsection{Different Knowledge of User Queries}\label{sec:query_distri}
We evaluate generalizability of our adversarial hubs across different query distributions by using text queries from other datasets (CUB-200, AudioCaps, MSR-VTT) to generate hubs for the text queries from MS COCO. \Cref{tab:different-distribution} shows that the attack remains effective for \emph{universal} adversarial hubs.  MSR-VTT yields the strongest transfer performance ($R@1=35.8\%, R@10=76.8\%$), likely because its visual-linguistic properties are similar to MS COCO and the attack only needs to approximate common visual concepts (e.g., objects and scenes) rather than match the exact query distribution.  Hubs optimized for CUB-200 exhibit the weakest transfer performance ($R@1=4.8\%$), which is expected given this dataset focuses on fine-grained object categories rather than descriptive captions. AudioCaps results are in the middle ($R@1=10.6\%$), showing that our attack exploits cross-modal semantics despite large modality differences.

In addition to the universal hubs, we evaluate \emph{concept-specific} hubs under the proxy knowledge setting, where MS~COCO serves as the target user distribution and a separate dataset provides the proxy queries. We define these concepts in two ways: word-based and cluster-based. For \emph{word-based} hubs, we identify words shared between the proxy and target (MS~COCO) captions. We compute the centroid using only the proxy text embeddings that contain a shared word, and evaluate the resulting hub strictly against the MS~COCO queries containing that same word. For \emph{cluster-based} hubs, we fuse the proxy and MS~COCO text embeddings and apply $k$-means clustering ($k{=}300$). For each mixed cluster, we compute the centroid using only its proxy members to generate the hub, and evaluate it against the MS~COCO queries within that same cluster. In both cases, the generated hub is appended to the MS~COCO gallery, and retrieval metrics are computed on the concept-matched MS~COCO query subset.

As shown in \Cref{tab:different-distribution}, concept-specific hubs significantly outperform universal hubs when semantic alignment exists. In particular, cluster-based hubs consistently achieve the strongest performance across all datasets (e.g., $R@1=96.7\%$ on AudioCaps and $91.1\%$ on MSR-VTT), indicating that aligning with the high-level semantic structure in the embedding space is more effective than relying on lexical overlap.  This is consistent with the exact-knowledge setting (\Cref{tab:concept}).  These findings indicate that our approach is robust to differences and shifts in query distributions and the attack exploits both localized regions of semantic similarity and global distributional alignment.

\subsection{Ablation Studies}\label{sec:ablation}
We conduct additional ablation studies to assess the impact of the target query distribution size ($|Q_t|$) and the perturbation budget ($\epsilon$) on adversarial hub performance in text-to-image retrieval tasks on the MS COCO dataset.

\begin{figure}[tb]
   \includegraphics[width=1.0\linewidth]{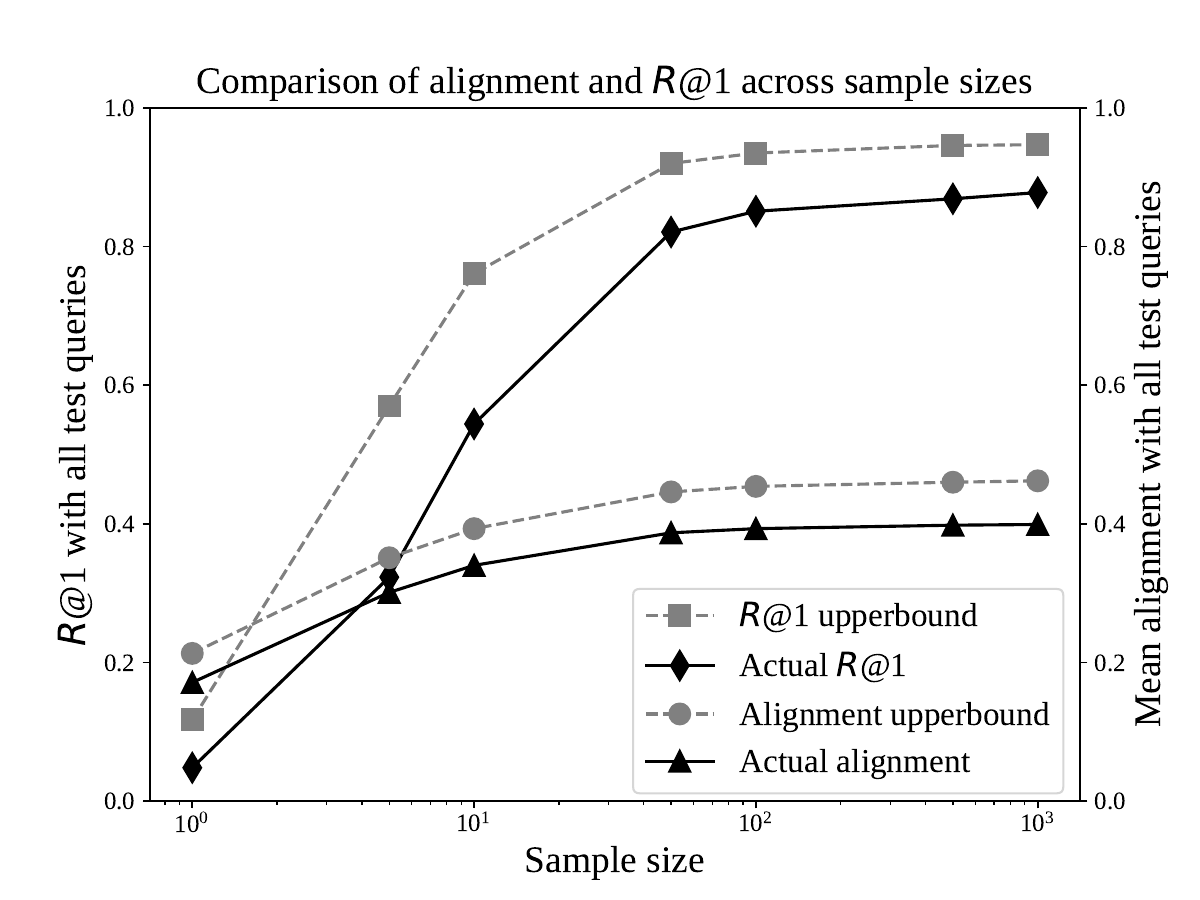}
   \centering
   \caption{\textbf{Performance vs. sample size.} Attack success rate (ASR) and cosine similarity between adversarial hub and query embeddings improve with larger target sample sizes, converging at around 100 samples.}
   \label{fig:sample_size}
\end{figure}

\para{Different Sizes of Target Query Set $Q_t$}
The $x$-axis in \Cref{fig:sample_size} denotes different sizes of the target query set $Q_t$. The default size of $Q_t$ used in our main experiments (\Cref{sec:wo_def_eval}–\Cref{sec:relaxed-assumptions}) is 100. The left $y$-axis shows the top-1 retrieval rate of adversarial hubs, each generated using a different $Q_t$, evaluated against the 25,000 test queries. The right $y$-axis shows the average cosine similarity between test query embeddings and their adversarial hubs.

The solid lines in the figure represent the performance of our optimized attack based on \Cref{eq:optimization_centroid} across varying $|Q_t|$. The dotted lines indicate the performance of a ``hypothetical'' optimal attack, where the adversarial hub embedding exactly matches the centroid $\mathbf{c}_t$ of $Q_t$. This upper-bounds both top-1 recall and average cosine similarity, since actual attacks may not align perfectly with $\mathbf{c}_t$.

From the figure, we observe that for both metrics, the solid and dotted lines increase with larger $|Q_t|$. However, the top-1 recall saturates around 85.1\% when $Q_t$ contains just 100 samples. Further increasing $|Q_t|$ yields only marginal gains. This suggests that adversarial hubs can be effectively generated using a relatively small set of queries (e.g., 0.4\% of the 25,000 total queries), while still closely approximating the embedding distribution of a much larger set. The diminishing returns beyond this saturation point are likely due to the compactness of the query embedding space, allowing 100 samples to provide a strong approximation, which our optimization method (\Cref{eq:optimization_centroid}) effectively exploits.

\para{Different Perturbation Bound}
\Cref{tab:different-eps} shows how the perturbation bound $\epsilon$ influences attack performance under the optimization in \Cref{eq:optimization_centroid}. Even small perturbations ($\epsilon = 8/255$) are effective, with 62.8\% $R@1$. Our default setting ($\epsilon = 16/255$), commonly used for generating adversarial examples, achieves 85.1\% $R@1$ with MdR = 1.0, striking a strong balance between attack success and imperceptibility. Larger bounds ($\epsilon = 32/255$, $64/255$) yield only marginal gains (90.5\% and 91.5\% $R@1$, respectively), indicating diminishing returns. This suggests that increasing the perturbation magnitude primarily raises detection risk without meaningful performance improvement, validating our chosen $\epsilon$ as an effective trade-off between effectiveness and stealthiness.

\renewcommand{\arraystretch}{1.2}
\begin{table}[tb]
  \caption{\textbf{Impact of perturbation bound $\epsilon$ on attack performance.} Results show how different perturbation magnitudes affect the retrieval metrics for adversarial hubs generated with ImageBind on MS COCO dataset.}
  \label{tab:different-eps}
  \centering
  \begin{tabular}{lrrrr}
     \toprule  
      {$\mathbf{\epsilon}$} & {\bf {\it R}@1(\%)}$\uparrow$ & {\bf {\it R}@5(\%)}$\uparrow$ & {\bf {\it R}@10(\%)}$\uparrow$ & {\bf MdR}$\downarrow$ \\ 
     \midrule
     $8/255$ & 62.8$_{\pm10.7}$ & 92.9$_{\pm4.4}$ & 96.4$_{\pm2.4}$ & 1.1$_{\pm0.4}$ \\
     $16/255$ & 85.1$_{\pm3.1{\color{white}0}}$ & 98.1$_{\pm0.7}$ & 99.0$_{\pm0.4}$ & 1.0$_{\pm0.0}$ \\
     $32/255$ & 90.5$_{\pm1.4{\color{white}0}}$ & 98.8$_{\pm0.3}$ & 99.4$_{\pm0.2}$ & 1.0$_{\pm0.0}$ \\
     $64/255$ & 91.5$_{\pm1.1{\color{white}0}}$ & 99.0$_{\pm0.3}$ & 99.5$_{\pm0.2}$ & 1.0$_{\pm0.0}$ \\
    \bottomrule
  \end{tabular}
\end{table}

\renewcommand{\arraystretch}{1.2}
\begin{table}[tb]
  \caption{\textbf{Impact of different geometric centers on attack performance.} Results show how different choices of geometric center affect the retrieval metrics for adversarial hubs generated with ImageBind on the MS COCO dataset.}
  \label{tab:different-centers}
  \centering
  \setlength{\tabcolsep}{5pt}
  \begin{tabular}{lrrrr}
     \toprule
      {\bf Center} & {\bf {\it R}@1(\%)}$\uparrow$ & {\bf {\it R}@5(\%)}$\uparrow$ & {\bf {\it R}@10(\%)}$\uparrow$ & {\bf MdR}$\downarrow$ \\
     \midrule
     Centroid         & 85.1$_{\pm3.1}$ & 98.1$_{\pm0.7}$ & 99.0$_{\pm0.4}$ & 1.0$_{\pm0.0}$ \\
     Trimmed centroid & 84.0$_{\pm3.6}$ & 97.8$_{\pm0.8}$ & 98.8$_{\pm0.5}$ & 1.0$_{\pm0.0}$ \\
     Medoid           & 11.8$_{\pm2.1}$ & 22.7$_{\pm5.2}$ & 29.2$_{\pm6.6}$ & 44.2$_{\pm16.7}$ \\
     Geometric median & 85.3$_{\pm3.5}$ & 98.1$_{\pm0.6}$ & 99.0$_{\pm0.4}$ & 1.0$_{\pm0.0}$ \\
    \bottomrule
  \end{tabular}
\end{table}

\para{Different Geometric Centers}
\label{different_metrics}
We use the centroid as the default geometric center because it is the most natural and intuitive proxy for a hub in embedding space. However, other geometric centers are also possible alternatives. To evaluate how sensitive our attack is to this design choice, \Cref{tab:different-centers} compares four options for constructing adversarial hubs: centroid, trimmed centroid, medoid, and geometric median. The results show that the centroid, trimmed centroid, and geometric median achieve very similar performance on MS COCO with ImageBind embeddings. Their $R@1$ values are 85.1\%, 84.0\%, and 85.3\%, respectively, and all three also achieve near-perfect $R@5$, $R@10$, and MdR. In contrast, the medoid performs substantially worse, with only 11.8\% $R@1$ and a much higher median rank. Overall, these results suggest that our attack is not tied to a specific notion of geometric center (as a proxy hub), although the centroid remains a simple and effective default.

\section{Related Work}\label{sec:related}

\para{Adversarial Alignment and Adversarial Examples}
The most relevant related work investigates \textit{adversarial alignment} of inputs in multi-modal embedding spaces~\cite{zhang2024adversarial, dou2024adversarial}.  These attacks incrementally adjust an input to align with a \textit{single} target embedding in an arbitrary modality.  By contrast, adversarial hubs achieve the much harder goal of aligning a single input to a \textit{large number} of target embeddings in arbitrary modalities.

Our work builds on the broader literature on untargeted and targeted adversarial examples.  Untargeted attacks cause outputs that are incorrect but not controlled by the adversary~\cite{goodfellow2014explaining}, while targeted attacks induce a specific, adversary-chosen, incorrect output~\cite{carlini2018audio}. By contrast, our approach generates adversarial inputs that work for \textit{multiple targets at once} (dozens of thousands of queries in the case of multi-modal retrieval).

In multi-modal settings, the most closely related papers are adversarial cross-modal examples~\cite{wu2024adversarialattacksmultimodalagents, zhao2023evaluating, dong2023how, peri2024speechguardexploringadversarialrobustness, gao2024adversarial, shayegani2023plug}. As well as being task-specific, these techniques only attack single targets while we demonstrate generalization to entire target distributions (queries related to a semantic concept). In other recent work~\cite{carlini2023aligned, qi2023visual,zhang2024soft}, untargeted adversarial perturbations are used for jailbreaking and prompt injection in multi-modal chatbots.  These problem settings are different from ours.

Hoedt et al.~\cite{hoedt2022defending} generate uni-modal audio adversarial hubs.  Their attack has two key limitations: it is (1) incompatible with modern retrieval systems based on neural encoders, and (2) unrealistically assumes the attacker has prior access to natural hubs.  By contrast, our attack converts \emph{any} data of the attacker's choice (in the same or different modality) into a hub using only a few random user queries and is naturally compatible with neural encoders.  Furthermore, our adversarial hubs impact a much larger number of queries, demonstrating a more scalable and realistic threat.  Other related work includes uni-modal adversarial examples against specific downstream tasks: toxicity classification~\cite{hosseini2017deceiving}, reading comprehension~\cite{jia2017adversarial}, and chatbots~\cite{zou2023universal}. Untargeted adversarial perturbations against contrastively-trained encoders are demonstrated in \cite{yu2022adversarial, kim2020adversarial, zhou2023downstream}.

\para{Poisoning Retrieval-Augmented Generation} 
Our work is also closely related to retrieval-augmented generation poisoning, which involves an adversary who injects adversarial documents into a RAG system's corpus. In this setting, adversaries create documents that have two goals: (1) they are frequently retrieved (retrieval poisoning) and (2) they result in unsafe generations when retrieved (generation poisoning).

Chaudhari et al.~\cite{chaudhari2024phantom} study trigger-based retrieval poisoning: they create adversarial documents that are retrieved for any query containing a specific trigger word. Others~\cite{zhong2023poisoning, xue2024badrag, ben2025gasliteing} study triggerless retrieval poisoning attacks by clustering clean documents and generating an adversarial document for each cluster. Zhang et al.~\cite{zhang2024adversarial} propose a method to produce natural-looking text that works for both trigger-based and triggerless attacks. After they are retrieved, adversarial documents cause generative models to produce harmful content~\cite{chaudhari2024phantom, shafran2024machineragjammingretrievalaugmented}.

Our work differs from prior RAG poisoning along three key dimensions. First, we introduce a new optimization target: unlike prior work, which maximizes similarity to individual queries incurring high computational costs, we optimize directly in embedding space to target entire classes of queries through their geometric structure (e.g., cluster centroids). Second, our setting is multi-modal, where alignment between discrete text and continuous modalities (e.g., images) is looser (\Cref{sec:perf_adv_hub}), making it possible for a single adversarial hub to attract many queries across modalities; in contrast, text-only RAG poisoning often requires injecting many adversarial documents \cite{zhong2023poisoning}. Third, our optimization operates in continuous domains, avoiding artifacts like unnatural phrasing or elevated perplexity, making these attacks both more effective and harder to detect than text-based approaches using token-level modifications.

While our work is more closely related to poisoning retrieval, an adversary could also target generation \textit{in any modality} by including malicious generation targets in their query set $Q_t$. We leave this to future work.

\para{Mitigation of Natural Hubness} Since hubs occur organically, without adversarial manipulation, in high-dimensional distributions, prior work focused on mitigating \textit{natural hubness}. There are train-time~\cite{liu2020hal} and post-processing methods \cite{bogolin2022cross,schnitzer2012local,suzuki2012investigating,wang2023balance} for mitigating natural hubs.

Train-time methods rely on hubness-aware loss functions \cite{liu2020hal} which downweigh points that are close to multiple neighbors. While effective, these methods are costly for large models and have been mostly supplanted by post-processing methods, which rescale abnormally high similarities between points. Early methods included combinations of local scaling~\cite{jegou2007contextual,zelnik2004self,schnitzer2012local}, global scaling~\cite{schnitzer2012local,hoedt2022defending}, and centroid-scaling~\cite{suzuki2012investigating,suzuki2013centering,hara2015localized}, but these are inefficient for large models.
More scalable methods leverage a query bank to normalize abnormal retrieval similarities \cite{bogolin2022cross,dinu2014improving,smith2017offline,conneau2017word,wang2023balance,chowdhury2024nearest}. Some are explicitly designed for cross-modal retrieval \cite{bogolin2022cross,wang2023balance,chowdhury2024nearest}.  We evaluate their effectiveness
in~\Cref{sec:hub_defense}.

\section{Conclusion and Future Work}
This paper investigated how the well-known phenomenon of hubness, where a data point in a high-dimensional space becomes a neighbor to many semantically unrelated points, can be adversarially exploited.  We focused this investigation on multi-modal, embedding-based retrieval methods that are at the heart of modern retrieval systems.

We introduced two types of adversarial hubs: (1) universal hubs that are close to, and thus retrieved by, many irrelevant user queries, and (2) concept-specific hubs that are only retrieved by queries related to adversary-chosen semantic concepts.  Through empirical evaluation, we showed that adversarial hubs can be crafted to appear relevant to many more queries than natural hubs.   Adversarial hubs could thus be (ab)used for spam and product promotion attacks on modern multi-modal retrieval systems that rely on pretrained embeddings.  Furthermore, we showed that defenses designed to mitigate natural hubness are ineffective against concept-based adversarial hubs, motivating research on new defenses.

There are several promising directions for future exploration. On the attack side, one could apply localized perturbations, akin to adversarial patches~\cite{brown2017adversarial}, to generate adversarial hubs, and extend their construction to modalities beyond images and audio, which are the focus of this paper. Another direction is to study adversarial hubs in the context of multi-modal RAG systems. On the defense side, mitigating adversarial hubness is a challenge, as demonstrated in this work. Future research may focus on designing efficient detection mechanisms against adaptive attacks and investigating tradeoffs between robustness and retrieval performance in multi-modal embedding models.

\section*{Ethical Considerations}
While our work investigates the vulnerability of multi-modal retrieval systems to adversarial hubs, our primary goal is to raise awareness of these threats and to inspire the development of effective defenses against adversarial hubs.

\section*{LLM Usage Considerations}

\noindent \textbf{Originality.} We used LLMs for light editing and to improve the flow of the text. We reviewed every change to ensure accuracy and originality. All technical contributions, including the definitions, attack designs, and proofs, were developed and written by the authors.

\noindent \textbf{Transparency.} LLMs were not used as part of our research methodology. They did not play a role in designing the attacks, implementing the systems, or analyzing the results. The reproducibility of our findings does not depend on any specific LLM service.

\noindent \textbf{Responsibility.} Our use of LLMs was limited to minor text edits and small coding support. No sensitive or proprietary data was shared with these models, and no new models were trained. The authors remain fully responsible for the entire content of the manuscript.

\section*{Acknowledgments}
Supported in part by an Amazon Research Award, Google
Academic Research Award, Google Cyber NYC Institutional
Research Program, a research gift from Infosys, and NSF awards 2311521 and 2428949.

\bibliographystyle{IEEEtranS}
\bibliography{reference}


\end{document}